\documentclass[aps,twocolumn,preprintnumbers,prd,superscriptaddress,nofootinbib,10pt]{revtex4-2}

\usepackage{amssymb,amsmath,amsthm,amsfonts}
\usepackage[scr=boondox]{mathalpha}
\usepackage[dvipsnames]{xcolor}
\definecolor{darkred}{rgb}{0.5,0,0}
\definecolor{linkcolor}{rgb}{0.0,0.3,0.5}
\usepackage[normalem]{ulem}
\usepackage{bm}

\usepackage{tensor,multirow}
\usepackage[
    colorlinks=true,
    citecolor=linkcolor,
    linkcolor=blue,
    filecolor=magenta,      
    urlcolor=cyan,
    pdfpagemode=FullScreen,
    linktocpage]{hyperref}
\usepackage{float}
\usepackage{graphicx}
\usepackage{physics}
\usepackage{siunitx}

\providecommand*{\iu}{\mathrm{i}\mkern1mu}                  %imaginary i
\newcommand{\pbilby}{\texttt{Parallel Bilby}}
\newcommand{\dynesty}{\texttt{Dynesty}}
\newcommand{\nrsur}{\texttt{NRSur7dq4}}

\newcommand{\ie}{{\it i.e.}}
\newcommand{\eg}{{\it e.g.}}

\def\<#1>{\mathinner{\langle#1\rangle}}
\newcommand{\lm}{\ensuremath{{\ell,m}}}
\newcommand{\lmm}{\ensuremath{{\ell,-m}}}
\newcommand{\vgw}{\ensuremath{V_{\rm GW}}}

\DeclareMathOperator\sgn{sgn}

\graphicspath{{./figures/}}

\begin{document}
\title{
    Gravitational-wave signatures of mirror (a)symmetry in binary black hole mergers: \\
    Measurability and correlation to gravitational-wave recoil.
}

 \author{Samson H.\,W. Leong}
    \email{samson.leong@link.cuhk.edu.hk}
    \affiliation{Department of Physics, The Chinese University of Hong Kong, Shatin, N.T., Hong Kong}

\author{Alejandro Florido Tom\'e}
    \email{afloto@alumni.uv.es }
    \affiliation{Departamento de Astronom\'{i}a y Astrof\'{i}sica, Universitat de Val\`{e}ncia, Dr. Moliner 50, 46100, Burjassot (Val\`{e}ncia), Spain}

\author{Juan~Calder\'on~Bustillo}
    \affiliation{Instituto Galego de F\'{i}sica de Altas Enerx\'{i}as, Universidade de Santiago de Compostela, 15782 Santiago de Compostela, Galicia, Spain}
    \affiliation{Department of Physics, The Chinese University of Hong Kong, Shatin, N.T., Hong Kong}

\author{Adri\'an del R\'io}
    \affiliation{Departamento de Matema\'ticas, Universidad Carlos III de Madrid, Avda. de la Universidad 30, 28911 Legane\'s, Spain.}

\author{Nicolas Sanchis-Gual}
    \affiliation{Departamento de Astronom\'{i}a y Astrof\'{i}sica, Universitat de Val\`{e}ncia, Dr. Moliner 50, 46100, Burjassot (Val\`{e}ncia), Spain}

\begin{abstract}
Precessing binary black-hole mergers can produce a net flux of circularly-polarized gravitational waves.
This imbalance between left- and right-handed circularly polarized waves, quantified via the Stokes pseudo-scalar $V_{\rm GW}$, originated from mirror asymmetries in the binary.
We scan the parameter space of black-hole mergers to investigate correlations between $V_{\rm GW}$ and chiral magnitudes constructed out of the intrinsic parameters of the binary. To this end, we use both numerical-relativity simulations for {quasi-circular} and eccentric precessing mergers from both the SXS and RIT catalogues, as well as the state-of-the-art surrogate model for quasi-circular precessing mergers NRSur7dq4. We find that, despite being computed by manifestly different formulas, $V_{\rm GW}$ is linearly correlated to the helicity of the final black hole, defined as the projection of its recoil velocity onto its spin. Next, we test our ability to perform accurate measurements of $V_{\rm GW}$ in gravitational-wave observations through the injection and recovery of numerically simulated signals. We show that $V_{\rm GW}$ can be estimated unbiasedly using the surrogate waveform model NRSur7dq4 even for signal-to-noise ratios of nearly 50, way beyond current gravitational-wave observations. 
\end{abstract}

\maketitle

\section{Introduction}
The detection of gravitational waves~(GWs) by the LIGO, Virgo and KAGRA interferometers~\cite{LIGOScientific:2014pky,VIRGO:2014yos,KAGRA:2018plz} provides accurate information on the dynamics of massive astrophysical compact objects, such as neutron stars or black holes~(BHs). The analysis of this information is particularly useful to test the non-linear aspects of General Relativity~(GR), and has provided numerous applications in astrophysics and cosmology~\cite{LIGOScientific:2018mvr,LVK2021:GWTC2,LVK2023:GWTC3,LIGOScientific:2019fpa,Populations_GWTC3, Vijay_GWKick,Hannam_nature_precession,Kick_GW190412,Abbott:2016blz, TGR_GWTC1, TGR_GWTC2, TGR_GWTC3,Isi2019_nohair,Bustillo2021,Siegel2023,H0_nature_lvk, Abbott_2021}. 

Binary black holes~(BBHs) constitute the most abundant astrophysical sources of gravitational waves~\cite{Populations_GWTC3}. These systems are characterized by 8 intrinsic parameters, namely the 2 masses and the 6 components of the two spin vectors, the distributions of which are linked to different astrophysical formation channels~\cite{Mapelli2020_Summary}. In particular, isolated binary evolution scenarios for star evolution favour BBH spin configurations in which the two vector spins are aligned, not only among themselves but also with the orbital angular momentum of the system~\cite{Tutukov1993, Belczynski2016}.  In contrast, dynamical formation channels, in which a binary is formed by gravitational capture of a BH by another one, favour isotropic spin distributions~\cite{Sigurdsson1993, PortegiesZwart2000, Rodriguez2016}. Therefore, GW observations have the potential to constrain these astrophysical formation channels with current and future data. 

It is easy to see that if the two spin vectors of a BBH remain aligned during the entire evolution, then the exchange of the two black holes in the binary at any instant of time renders exactly the same binary system. This is because the BBH always remains invariant under a mirror transformation with respect to the separating plane~\cite{dRetal20, sanchis2023precessing}. However, if the two spin vectors are misaligned between themselves and with respect to the orbital angular momentum of the binary, the system will fail to be invariant under such a mirror transformation. We can thus use mirror (a)symmetry to study the relative spin configuration of the two BHs in a binary, as well as to gain further insights into the dynamics of BBHs~\cite{sanchis2023precessing}. 

As we will explain below in more detail, mirror asymmetry leaves a characteristic imprint on the GW emission of the BBH. Namely, if the BBH fails to be invariant under mirror transformation, then the net GW emission over its full celestial sphere is circularly polarized. It is easy to see why this has to be the case. Under a mirror transformation, left-handed GW modes turn into right-handed modes, and vice versa. If there is an excess of one handedness over the other, then the total emitted flux of GW will fail to be invariant under a mirror transformation. Therefore, a mirror-symmetric BBH must necessarily produce equal amounts of right- vs left-handed GW modes. 

The above feature has been used to test the Cosmological Principle. Individual precessing BBHs are mirror asymmetric, therefore producing a net flux of circularly polarized waves. However, if the Cosmological Principle holds, the combined emission of all sources should average out, yielding the same amount of left- and right-circularly polarized waves. In a recent work, we showed that the gravitational-wave event GW200129~\cite{LVK2023:GWTC3}, which displays orbital precession~\cite{Hannam_nature_precession} is indeed mirror asymmetric (see also: Refs~\cite{Payne2022:GW200129_DQ,Macas2023:GW200129_ML}). However, using an ensemble of 47 BBHs analyzed by Islam {\it et al.}~\cite{Islam:2023zzj} we showed that, on average, they produced a net flux of circularly polarized waves $\langle \vgw\rangle$ consistent with zero, therefore respecting the Cosmological Principle~\cite{toappear}.

This work has two main goals. First, we aim to find a direct connection between the net circular polarization of GWs, quantified with the observable $\vgw$ in Eq.~\eqref{eq:VGW} below, with the intrinsic properties of the source BBH. As we will describe in detail below, $\vgw$ is an intrinsic pseudo-scalar of the system that is computed through the integration of gravitational-wave fluxes throughout the whole evolution of the system. Because of this, we study correlations between $\vgw$ and chiral magnitudes of the  BBH. Remarkably, we find that, for BBHs, $\vgw$ is linearly correlated with the projection of the recoil of the final black hole onto its spin (\ie\ its helicity). Second, we study our ability to correctly estimate $\vgw$ in current and future gravitational-wave observations of BBHs. We do this by performing parameter inference on synthetic numerical-relativity waveforms injected in gravitational-wave detectors, using the state-of-the-art surrogate model \nrsur~\cite{NRSur7dq4}. We find that $\vgw$ can be measured in an unbiased way for signal-to-noise ratios of 50, way above those of current gravitational-wave observations.

The rest of this article is organized as follows. In Sec.~\ref{sec:MirrorAsym}, we review the basic framework on mirror (a)symmetry in BBHs, and its connection with the gravitational Stokes $\vgw$ parameter, which measures the emission of net GW circular polarization, or GW helicity. In Sec.~\ref{sec:PE} we analyze our ability to measure $\vgw$ from GW observations. Next, in Sec.~\ref{sec:BHrecoil_and_VGW} we describe linear correlations between $\vgw$ and the projection of the final recoil onto the final spin and the total angular momentum measured close to the merger. Finally, we discuss our results in Sec.~\ref{sec:Discussion}.

We use geometrized units throughout this paper, \ie~$G=c=1$. We work with effective spatial vectors in $\mathbb R^3$, as usual in the Post-Newtonian framework. In particular, 3-dimensional vectors are represented by bold-face symbols, carets denote unit vectors, and the absence of bold-face denotes the corresponding magnitudes, \eg\ $K$ denotes the magnitude of $\vb*K$. On the other hand, we follow the notation of Ref.~\cite{wald:1984} for the metric signature and Riemann tensor.  

\section{Mirror asymmetry and GW circular polarization}\label{sec:MirrorAsym}

In this section, we review the framework of mirror (a)symmetry developed in~\cite{dRetal20, sanchis2023precessing} for BBHs and explain its connection with net circular polarization in the emitted GW signal.

\begin{figure}[t!]
    \centering
    \includegraphics[width=1.0\linewidth]{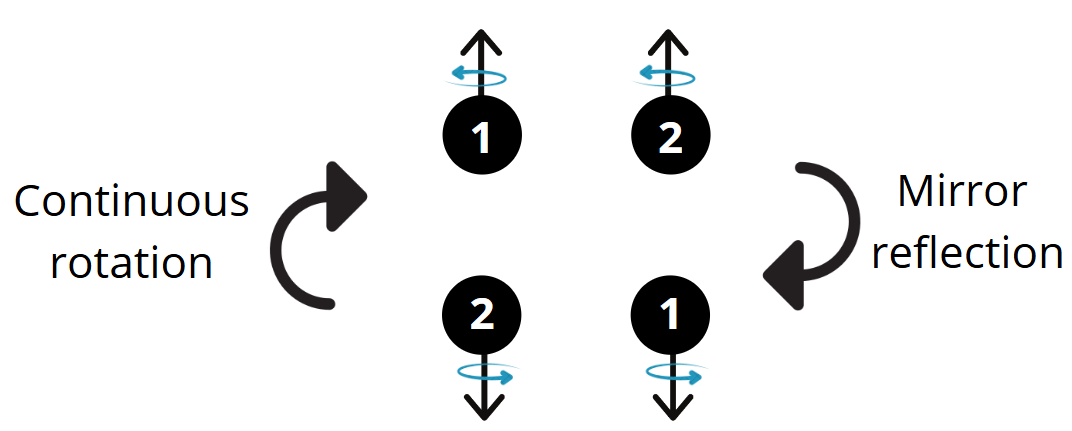}
    \caption{
        Example of a BBH that remains invariant under a mirror transformation with respect to the separating plane. The picture represents one instant of time during the inspiral. The arrows denote the individual spin vectors of the two BHs, each one labelled as 1 and 2. As a consequence of the spacetime mirror symmetry, the Stokes parameter (Eq.~\eqref{eq:VGW}) vanish~\cite{dRetal20, sanchis2023precessing}. 
    }
    \label{fig:diagram1}
\end{figure}

To illustrate the notion of mirror symmetry in a BBH it is convenient first to analyze a system in which the two spin vectors are parallel to the orbital angular momentum, as displayed in Fig.~\ref{fig:diagram1}. For conceptual simplicity, let us assume that the two black holes are sufficiently separated, so that non-linearities are negligible and we can focus on the properties of the two individual Kerr BHs.
This approximation is reasonable during the inspiral phase, as black holes have zero or very small tidal Love numbers and are not significantly deformed~\cite{Perry2023:DynamicalLove}. In this crude approximation, the spacetime of the binary can be fully described by the two masses and the two spin vectors. 

If we fix an arbitrary instant of time during the inspiral, the BBH appears as in the upper figure of Fig.~\ref{fig:diagram1}.  As one can notice, the picture displays a manifest mirror symmetry. Namely, {in this instant of time we can introduce a coordinate system where the binary is separated \eg~along the $x$ is the axis, and we can naturally define a separating plane between the two BHs as the coordinate plane $\{x=x_0\}$ that divides the $x$-distance among the two.} A mirror reflection with respect to the separating plane renders the lower panel of Fig.~\ref{fig:diagram1}. This can be easily verified by recalling that the spins and orbital angular momentum are pseudo-vectors, \ie\ they flip sign under an improper transformation. Given this mirror-transformed system, it is easy to visualize that there exists a continuous spatial rotation that returns the BBH back to its original configuration of spins and masses, so the original system remains invariant. 
Because the two spins remain aligned with the orbital angular momentum, the system will not precess in time. In particular, the orientation of the orbital angular momentum will remain constant during the entire evolution of the BBH. The analysis above can therefore be applied at any instant of time. This is, we do not expect such a BBH to produce any spacetime mirror asymmetry during its evolution. In fact, numerical simulations confirm the prediction that non-precessing systems do not produce mirror asymmetry for a wide range of BBH parameters~\cite{sanchis2023precessing}. 

\begin{figure}[t!]
    \centering
    \includegraphics[width=1.0\linewidth]{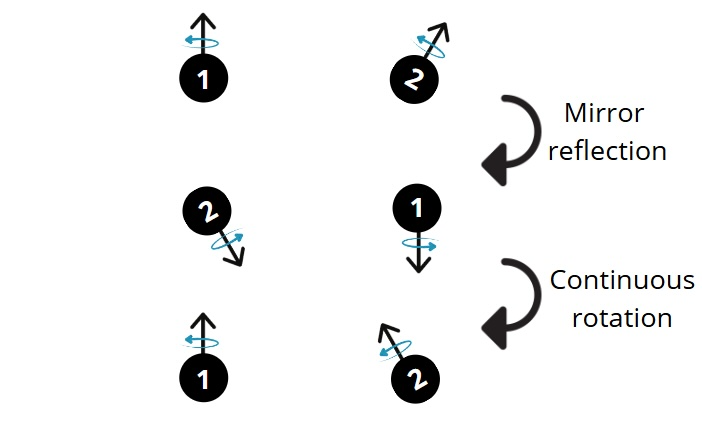}
    \caption{
        Example of a BBH that fails to be invariant under mirror transformations. The lack of spacetime mirror symmetry, produced as a consequence of the misalignment of the two BH spin vectors, leads to Stokes parameter (Eq.~\eqref{eq:VGW}) that differ from zero. 
    }
    \label{fig:diagram2}
\end{figure}

However, if the two spin vectors in the BBH fail to be aligned, the mirror symmetry in the preceding example would be spoiled. An example of this is shown in Fig.~\ref{fig:diagram2}, where the two BHs are displayed at a fixed instance of time.
In sharp contrast to the previous example, in this case, a mirror transformation with respect to the separating plane renders a new BBH system that cannot be transformed back to the original system by a continuous spatial rotation.
The above implies that mirror asymmetry requires the system to display misaligned spins, which in turn leads to orbital precession. We note that the converse is not necessarily true, as emissions from certain precessing configurations may average out, yielding a null $\vgw$.

The qualitative analysis presented above calls for a mathematically precise definition of mirror (a)symmetry, that will allow us to work on a rigorous footing, and to generalize this qualitative notion beyond BBHs, which may not be easily visualized. 
We will focus on astrophysical systems, for which any matter distribution is confined in a bounded region of space for any instant of time. These systems can be accurately described with the mathematical framework of asymptotically flat spacetimes~\cite{Geroch1977, Ashtekar:1987tt, AshtekarAbhay2015Gapo}. 

Given an asymptotically flat spacetime $({\cal M},g_{ab})$ with manifold ${\cal M}$ and metric $g_{ab}$, we can assign a precise notion of mirror asymmetry by calculating the integral~\cite{dR21}:
\begin{equation}
    \vgw := \int_{\mathcal J^+} \dd \mathcal J^+ \epsilon^{abc}N_{ad}D_b {N_c}^d \ ,
    \label{eq:chernsimons}
\end{equation}
where $\mathcal J^+$ represents future null infinity (a 3-dimensional null hypersurface attached to our manifold ${\cal M}$, see \eg\ Refs.~\cite{Geroch1977, Ashtekar:1987tt, AshtekarAbhay2015Gapo}), $N_{ab}$ denotes the Bondi News tensor, $D_a$ is a canonical covariant derivative on $\mathcal J^+$~\cite{PhysRevLett.46.573}, and $\dd {\cal J}^+\epsilon^{abc}$ is the 3-volume form on $\mathcal J^+$.

There are several reasons that make Eq.~\eqref{eq:chernsimons} a suitable physical observable to quantify the degree of mirror symmetry of $({\cal M},g_{ab})$. First of all, $\vgw$ is a (pseudo)scalar. In other words, it is invariant under any coordinate transformation that preserves parity.  It is therefore an intrinsic geometric parameter of the spacetime and can then be used to obtain intrinsic information of astrophysical sources. 
In particular, it is invariant under spatial rotations, hence independent of the orientation of the BBH with respect to the observer. 
In addition, Eq.~\eqref{eq:chernsimons} is {manifestly} chiral (as the presence of $\epsilon^{abc}$ indicates), flipping sign under those coordinate transformations that reverse the handedness of the frame, such as a parity or mirror transformation.
This implies that, if a given BBH remains invariant under a mirror transformation, we must necessarily have $\vgw=-\vgw$ and therefore $\vgw=0$. Since~\eqref{eq:chernsimons} is a dimensionless real number, we can therefore use this quantity to assign a notion of handedness to $({\cal M},g_{ab})$. 
We will say that $({\cal M},g_{ab})$ has positive handedness if $\vgw>0$, no handedness if $\vgw=0$, and negative handedness otherwise.  The higher the value of $\vgw$, the higher the degree of mirror asymmetry is in $({\cal M},g_{ab})$.

Numerical studies~\cite{dRetal20} have validated Eq.~\eqref{eq:chernsimons} as a suitable observable that reproduces all the points discussed in the qualitative analysis presented at the beginning of this section. It is therefore a potential candidate to quantify our intuitive notion of mirror (a)symmetry.
Eq.~\eqref{eq:chernsimons} has the form of a Chern-Simons current across future null infinity. 
In particular, it only depends on the asymptotic GW data. 
Indeed, it is well known that a non-trivial Bondi News tensor reveals the presence of a flux of GW across future null infinity~\cite{Geroch1977}. Thus, according to Eq.~\eqref{eq:chernsimons}, the handedness of our spacetime geometry can be identified with the emission of a chiral flux of GWs. Using the standard relation of $N_{ab}$ with the complex GW strain $h = h^+ - \iu h^\times$ in a generic frame, one can find a more explicit formula~\cite{dRetal20, dR21}:
\begin{equation}
\begin{split}
\vgw = &\int_{0}^{\infty} \dd \omega \,\omega^3 \sum_{\ell=0}^{\infty}\sum_{m=-\ell}^{+\ell} \\
       &\qty[ \abs{ \widetilde{h^+_\lm}(\omega) + \iu \widetilde{h^\times_\lm}(\omega) }^2 
   - \abs{ \widetilde{h^+_\lm}(\omega) - \iu \widetilde{h^\times_\lm}(\omega) }^2 ]\ ,
\end{split}
\raisetag{3.5\baselineskip}
\label{eq:VGW}
\end{equation}
where $h_\lm$ are the modes of the GW strain $h$ in the spin-weighted spherical harmonic basis (see Eq.~\eqref{eq:hlmYlm}), and $h^+_\lm = \Re[h_\lm]$ and $h^\times_\lm = -\Im[h_\lm]$, with the overhead tilde denotes the Fourier transform of them.
Physically, the combinations $\tilde h^+_\lm - \iu \tilde h^\times_\lm$, $\tilde h^+_\lm + \iu \tilde h^\times_\lm$, represent a left- and right-handed circularly polarized wave mode, respectively~\cite{Isi2023:Polarisation}.
Consequently, Eq.~\eqref{eq:VGW} measures the net circular polarization of the GW flux emitted by any isolated astrophysical sources. Because of this, we refer to $\vgw$ as the gravitational Stokes $V$ parameter, in analogy to the usual Stokes $V$ parameter for electromagnetic waves.

The formula above for $\vgw$ explicitly realizes the connection between mirror asymmetry and non-zero GW circular polarization. More precisely, an astrophysical system with mirror symmetry, such as the merger of two spinning black holes with parallel spins, always admits at least one coordinate frame where the following equalities hold (a consequence of reflection symmetry)
\begin{align}
    \tilde h^{+}_\lmm &= (-1)^{\ell} \overline{\tilde h^{+}_\lm}\, , \label{eq:m1}\\
    \tilde h^{\times}_\lmm &= -(-1)^{\ell} \overline{\tilde h^{\times}_\lm}\,  . \label{eq:m2}
\end{align}
Consequently, even if a particular GW mode $(\tilde h^{+}_{\ell m},\tilde h^{\times}_{\ell m})$ may be circularly polarized, in the sense that $|\tilde h^{+}_{\ell m}-i \tilde h^{\times}_{\ell m}|^{2}-|\tilde h^{+}_{\ell m}+i \tilde h_{\ell m}^{\times}|^{2}\neq 0$, the  mirror symmetric mode $(\tilde h^{+}_{\ell \, -m},\tilde h^{\times}_{\ell\, -m})$ will cancel this contribution out in \eqref{eq:VGW}. This is, mirror symmetry implies $\vgw=0$, \ie\ null net circular polarization, as expected on general grounds (as a matter of fact, $\vgw$ flips sign under a mirror transformation).  In order to get an imbalance between right-handed and left-handed GW waves, $\vgw\neq 0$, mirror symmetry must be broken

\section{Measuring $\vgw$ from gravitational-wave observations of black-hole mergers}
\label{sec:PE}

In this section, we test our ability to measure $\vgw$ from gravitational-wave observations. To this end, we inject  several numerically simulated signals from eight quasi-circular generically spinning black-hole mergers computed by the \texttt{SXS} collaboration~\cite{SXS} in a triple Advanced LIGO Hanford, Livingston and Virgo detector network. The parameters of the BBHs can be found in Table~\ref{tab:sxs_injections}. Next, we perform full Bayesian parameter inference on such signals using the \nrsur\ waveform model~\cite{NRSur7dq4}, which is directly fitted to numerical simulations of BBHs including orbital precession and higher-order modes. We perform our injections in zero-noise, using the design power-spectral densities of the three detectors~\cite{advLIGOcurves,advVIRGO}. For each of the simulated BBHs, we inject signals corresponding to three different source inclinations, scaling them to have signal-to-noise ratios of both $17$ (typical among current observations) and $50$, way beyond the loudest detection to date. We perform our parameter inference using the publicly available \pbilby~\cite{Ashton:2018jfp,pbilby} library. We sample the parameter space using the nested-sampling algorithm \dynesty~with 4096 live points, setting usual priors in all of the 15 source parameters. 

\begin{table*}[htb]
    \centering
    \begin{tabular}{c|cccccS[table-number-alignment = center]}
        \toprule
        SXS code & $Q = m_1 / m_2$ & $a_1$ & $a_2$ & $\chi_{\rm eff}$ & $\chi_{\rm p}$ & $\vgw$ \\ \hline
        \texttt{SXS:BBH:1443} & 5.681 & 0.4079 & 0.7372 & 0.2365 & 0.0000 &    0.000 \\
        \texttt{SXS:BBH:0045} & 3.000 & 0.4995 & 0.4994 & 0.2498 & 0.0000 &   -0.000 \\
        \texttt{SXS:BBH:0283} & 3.000 & 0.3000 & 0.2999 & 0.3000 & 0.0000 &    0.000 \\
        \texttt{SXS:BBH:1805} & 3.415 & 0.4849 & 0.7130 & 0.3743 & 0.2215 &   -0.474 \\
        \texttt{SXS:BBH:1676} & 3.253 & 0.4856 & 0.4018 & 0.3841 & 0.2245 &   -0.052 \\
        \texttt{SXS:BBH:1156} & 4.387 & 0.4663 & 0.7677 & 0.3299 & 0.2719 &   -0.052 \\
        \texttt{SXS:BBH:1410} & 4.000 & 0.4680 & 0.4647 & 0.2525 & 0.4000 &    0.137 \\
        \texttt{SXS:BBH:1593} & 3.500 & 0.7213 & 0.7588 & 0.2531 & 0.6866 &    1.028 \\
        \toprule
    \end{tabular}
    \caption{{\bf Parameter of the SXS numerical simulations of BBHs used for our injection study.} 
    The first column shows the SXS code of the NR simulation. 
    From the second column onwards, we show the mass ratio $Q$, dimensionless spin magnitudes $a_{1,2}$, effective spin parameter $\chi_{\rm eff}$ and effective precession spin parameter $\chi_{\rm p}$~\cite{chip_definition}, measured at the reference time of the simulation. 
    The last column shows the values of $\vgw$ computed from Eq.~\eqref{eq:vgw_psi4}.
    The table is ordered by increasing values of $\chi_{\rm p}$.
    }
    \label{tab:sxs_injections}
\end{table*}

\begin{figure*}
    \centering  
    \includegraphics[width=1\textwidth]{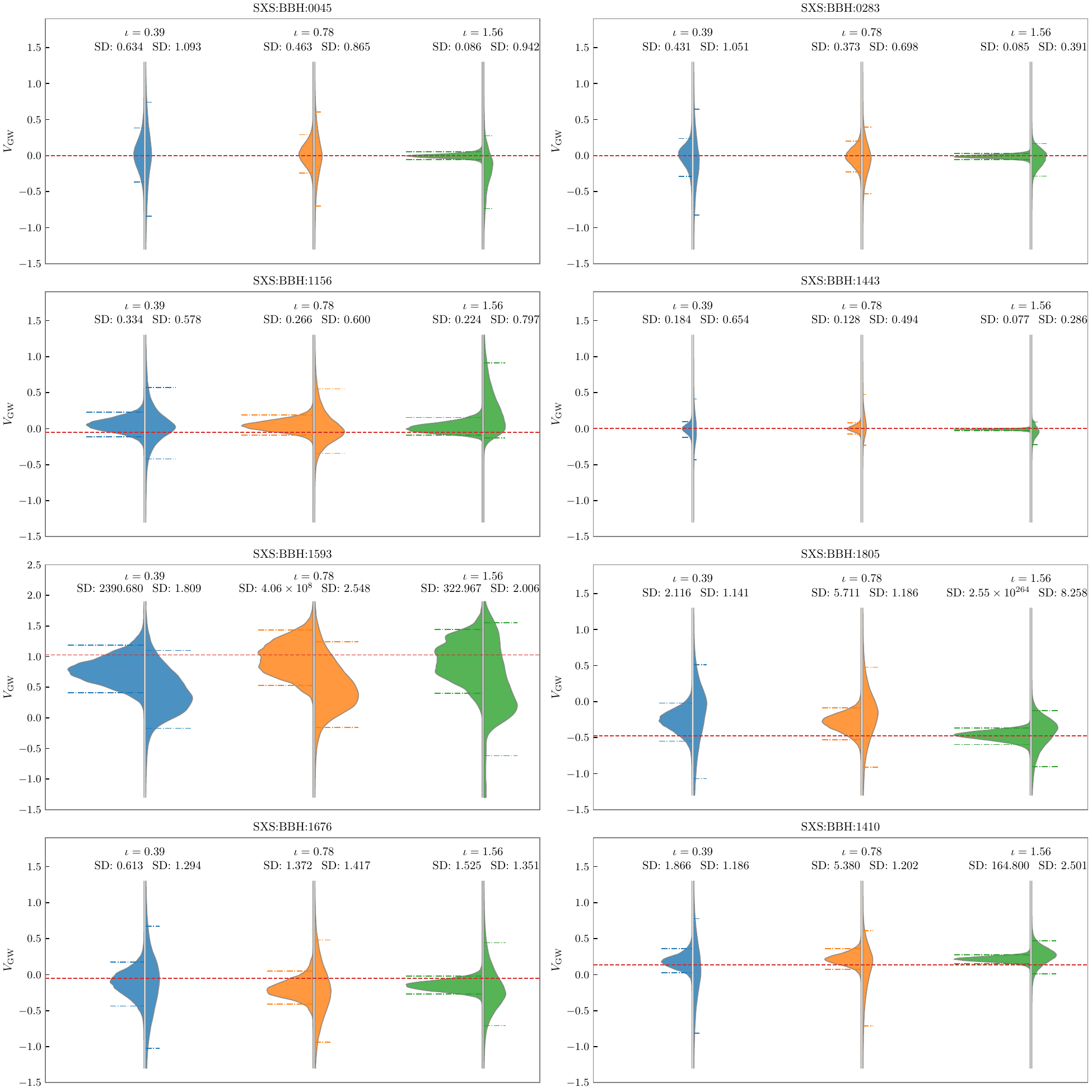}
    \caption{\textbf{Recovery of the GW Stokes $\vgw$ parameter for numerically simulated signals}. 
    Posterior distributions of the GW Stokes $\vgw$ parameter, obtained from the recovery of numerically simulated signals, for eight different BBHs observed at three inclinations, indicated at the top of each plot.
    The right (left) violins correspond to signal-to-noise ratios of 17 (50). 
    The coloured dashed lines on each distribution enclose the 90\% credible intervals, and the common red dashed line denotes the true $\vgw$ value computed from the simulation.  
    The acronym ``SD'' refers to the value of the Savage-Dickey ratio evaluated at $\vgw = 0$, as detailed in Eq.~\eqref{eq:SDratio}.
    }
    \label{fig:pe}
\end{figure*}

Figure~\ref{fig:pe} shows the posterior distributions for the $\vgw$ parameter. The right distributions correspond to injections with a network SNR of 17, while the left ones correspond to a network SNR of 50. As our main result, we note that in all cases the true injected value lies within the $90\%$ credible interval, even for the loudest cases, showing that $\vgw$ can be estimated in an unbiased way. Given its relation with the gravitational recoil, which we will discuss later, we understand that our results are consistent with those shown in the Supplementary Material of~\cite{Kick_GW190412}, which shows that accurate recoil measurements (in terms of magnitude and direction) can also be performed for SNRs of 50. The precision of the measurement generally increases for signals emitted at larger inclinations $\iota$, as the higher harmonics become more dominant~\cite{Blanchet:1995ez,Bustillo:2015ova} and allow to extract more detailed information about the source by \eg\ breaking parameter degeneracies~\cite{Kick_GW190412}. 

Next, we focus on model selection. This is, for each of our injections, we perform model selection between a ``mirror-symmetric'' model restricted to $\vgw=0$ and a generic model allowing for non-zero $\vgw$. To this, we compute the ratio of the Bayesian evidences ${\cal B}^{\vgw = 0}_{\vgw\neq 0}$ obtained for each of the two models. Following Ref~\cite{toappear}, this can be easily obtained through the Savage-Dickey ratio, {which is the ratio of the prior distribution $\pi(\vgw)$ and the posterior distribution $p(\vgw\mid d)$ at $\vgw = 0$, \ie}
\begin{equation}
    {\cal B}_{\vgw = 0}^{\vgw \neq 0} = {\rm SD} \equiv \frac{\pi(\vgw = 0)}{p(\vgw = 0 \mid d)}\, .
    \label{eq:SDratio}
\end{equation}
{where $d$ represents the GW strain data.}
For cases where the true $\vgw$ is null, we find that the Bayes' factor is either below or nearly one, indicating either a preference for $\vgw=0$ or lack of preference for any model. In particular, we note that such preference is stronger for near edge-on cases characterised by $\iota = 1.56$. 
Next, for cases with $\vgw \neq 0$ such as \texttt{SXS:BBH:1593}, \texttt{SXS:BBH:1805} and \texttt{SXS:BBH:1410}, the analyses returns preference for the correct $\vgw \neq 0$ model that becomes stronger with increasing SNR. In summary, for all cases, we find that the Bayes' factor either returns a strong preference for the correct model or remains rather inconclusive, preventing a biased model selection. 
{We note that Gaussian kernel density estimates (KDE) were used when evaluating the posterior value at $\vgw = 0$. The  extremely high Bayes' factors seen in some cases at SNR = 50 are highly sensitive to how the KDEs approximate locations where no samples is present. In our case, we employed the ``silverman'' bandwidth approximation implemented in \texttt{SciPy}~\cite{scipy2020}.}

\section{Black-hole recoils and $\vgw$}
\label{sec:BHrecoil_and_VGW}
Strong evidence for mirror asymmetry $\vgw\neq 0$ has only been found on one current GW event, namely GW200129~\cite{toappear}. Incidentally, this is also the only current GW event that has been claimed to display both orbital precession~\cite{Hannam_nature_precession}, together with a strong kick for the remnant BH directed out of the orbital plane~\cite{Vijay_GWKick}. In this section, we review the concept of black-hole recoil to then describe interesting correlations we have found between this observable and $\vgw$. 

Black-hole recoil is a strong-gravity effect sourced by the asymmetric emission of linear momentum by binary black-hole mergers in the form of gravitational waves~\cite{Thorne:1980ru,Gonzalez:2006md,Campanelli2007_kick}. As a result, the remnant black hole of a black-hole merger can acquire velocities that, depending on the binary parameters, can reach velocities up to $\SI{5000}{\km\per\s}$~\cite{Campanelli2007_kick,Sperhake:2010uv}, which are large enough to escape any host environment other than Active Galactic Nuclei~\cite{Baumgardt2018} (but see also Ref.~\cite{Gilbaum2024:EscapeAGN}). This has strong implications in black-hole formation channels, particularly impacting the viability of hierarchical black-hole formation~\cite{Gerosa2021,Arajolvarez2024,Mahapatra2021:Kick,Mahapatra2022:hierarchical}. Analytically, the magnitude and direction of the recoil is determined by the asymmetries of the system, which are encoded in the multipole structure, $h_\lm$ of the gravitational-wave emission~\cite{Ruiz2007:Multipole,Boyle2014:Precession}.

For non-precessing sources, which trivially satisfy mirror symmetry, positive and negative $m$ modes are related by $h_\lmm=(-1)^{\ell}\overline{h_\lm}$ in some frame, \eg~in that frame where the $z$ axis is aligned with the orbital angular momentum. In these cases, the kick arises from the interaction of multipoles of both even and odd parity (\ie, even and odd $m$), which occurs for unequal-mass sources and allows to define a preferred direction within the orbital plane~\cite{CalderonBustillo:2018zuq,CalderonBustillo:2019wwe,Kick_GW190412}. Once such asymmetry is present, the magnitude of the kick is then mostly determined by the net radiated energy, which increases towards equal-mass systems and positive aligned spins, {reaching values of up to \SI{500}{\km\per\s}~\cite{Healy2014_alignedspinkick}.} Precessing systems do not satisfy the above equation, which prevents the emissions from the co-dominant $(\ell,m)=(2, \pm 2)$ modes from cancelling each other. In this situation, the asymmetry between these modes can lead to much-increased kick magnitudes, reaching \SI{3000}{\km\per\second} in the so-called ``superkick'' configurations~\cite{Campanelli2007_kick,Sperhake:2010uv,Ma2021:Superkick,Mielke:2024kya}, which can be further increased on the presence of orbital eccentricity~\cite{Sperhake2020_ecc}.

The fact that both the kick and $\vgw$ are closely related to asymmetries in the multipole structure of the corresponding sources (see App.~\ref{app:asym_modes}), invites the question of whether these two observables are correlated and how. 
For instance, as argued in previous sections, orbital precession is required for producing a non-zero $\vgw$ (the converse is not necessarily true). At the same time, orbital precession is required for the remnant BH to acquire strong kicks and, moreover, for the latter to have a non-zero off-plane component, letting the remnant BH escape the orbital plane. On the other hand, strong precession (and therefore strong kicks) normally imply that the remnant black-hole recoils in a direction closely aligned (or anti-aligned) with the total angular momentum of the binary, and therefore with the final spin. This feature is also intuitively indicative of a strong imbalance between right and left-handed GW emissions.

\begin{figure}
    \centering
    \includegraphics[width=1.0\linewidth]{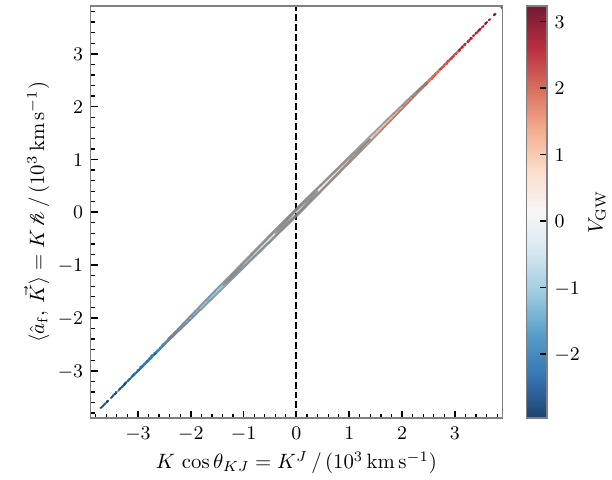}
    \caption{{\bf Linear correlation between the helicity $\mathscr{h}$, the transverse recoil velocity  $K^J$, and the gravitational Stokes parameter $\vgw$.}
    The $x$ and $y$ axes show, respectively, the projections of the recoil velocity $\vb*{K}$ onto the total angular momentum direction $\vu*{J}$ and onto the final BH spin direction $\vu*{a}_{\rm f}$. The latter quantity can be understood as proportional to the helicity $\mathscr{h}=\langle  \vu*{K}, \vu*{a}_{\rm f} \rangle$ of the remnant black hole. 
     The colour denotes the value of $\vgw$: red (blue) indicates positive (negative) values, which agree with the signs of $K^J$ and $\mathscr{h}$.
     The points correspond to an ensemble of \num{500000} BBHs drawn from a population with uniform masses and isotropic spins measured at a reference time $t_{\rm ref} = -100\,M$.
     All three quantities are computed from GW strains generated from the \nrsur\ model.
     The grey contours overlaying the points denote the (2D) $\{1,\ 2,\ 3\}\,\sigma$ levels of the distribution, representing the density of the population on this plane.
     }
    \label{fig:linear_relation}
\end{figure}

\subsection{Theoretical expectations\label{sec:theory_expectation}}

The potential connection between the kick and $\vgw$ can be first approached analytically. The latter is computed from Eq.~\eqref{eq:VGW}. 
On the other hand, in order to conserve the linear momentum of the system, the remnant BH recoils (or kicks) in the opposite direction to the net emission of linear momentum flux carried by GW~\cite{Peres1962:RadiatianRecoil}.
This can be computed through the direct integration of the linear momentum flux computed from the multipoles of the GW emission, $h_\lm$~\cite{RevModPhys.52.299, Gonzalez:2006md, Ruiz2007:Multipole, Boyle2014:Precession}.

To lay the ground for discussion, the complex GW strain $h$ can be expressed in terms of the spin\mbox{(-2)}-weighted spherical harmonics ${}_{-2}Y_\lm$~\cite{NP62,RevModPhys.52.299}:
\begin{align}
h(u,\theta,\phi) &= h^+(u,\theta,\phi) - \iu h^\times(u,\theta,\phi) \nonumber \\
                 &= \sum_{\ell=0}^{\infty}\sum_{m=-\ell}^{+\ell} h_\lm(u) \, {_{-2}Y}_\lm(\theta,\phi) \label{eq:hlmYlm} \\
                 &= \sum_{\ell=0}^{\infty}\sum_{m=-\ell}^{+\ell} (h^+_\lm(u)- \iu h^\times_\lm(u)) {_{-2}Y}_\lm(\theta,\phi) \nonumber
\end{align}
where $h^+$, $h^\times$ are the two real GW linear polarizations, $h_\lm$ denotes the multipoles of $h$ in this expansion, and can be separated in their real and imaginary parts, $h^+_\lm$ and $h^\times_\lm$ respectively. 
The retarded time is denoted by $u$, and the angles $(\theta,\,\phi)$ are the polar and azimuthal angles in a Cartesian frame adapted to the binary, in which the $z$-axis is parallel to the total angular momentum $\vb*J$, defined at some reference time during the evolution of the binary.
Then, following the notation in Ref.~\cite{Boyle2014:Precession}, the linear momentum flux in the direction $\vu*r$ can be expressed as:
\begin{equation}
    \frac{\dd \vb*P}{\dd\Omega\,\dd u} = \frac{1}{16\pi}\,\abs{\dv{h}{u} }^2 \,\vu*r \ ,
    \label{linearmomentum}
\end{equation}
where $\Omega$ and $u$ are the solid angle and retarded time respectively.
The net emitted linear momentum $\vb*P$ by the astrophysical source can be obtained upon integrating over all angles and time. 
In particular, the component along the $\vb*J$ direction, $P^J = \< \vb*P,\,\vu*J>$, is the $z$-direction in the chosen frame, and its leading order terms are found as:
\begin{equation}
     P^J = \int_{-\infty}^{\infty}\frac{\dd u}{{24\pi}} 
     \qty[ \abs{ \dot h^+_{2,2} -\iu  \dot h^\times_{2, 2}}^2
     - \abs{ \dot h^+_{2,-2} - \iu \dot h^\times_{2,-2} }^2 ] + \ldots 
     \label{eq:PJz_leading}
\end{equation}
where the overhead dot denotes derivative w.r.t.~retarded time $u$, and the ellipsis suppresses contributions from higher-order multipoles.
Then the recoil velocity (or kick) is defined as $\vb*K = - \vb*P / M$, with $M$ denoting the mass of the remnant BH, and similarly for $K^J$. 
By expanding now the multipoles $h_\lm^+$, $h_\lm^\times$ in Fourier modes, the leading order contribution can be expressed as\footnote{We have tacitly suppressed the subdominant contributions other than the $ (2,\,\pm 2)$ modes in the following expressions.}: 
\begin{alignat}{2}
    K^J &\approx - \frac{1}{M}&& \int_0^\infty \frac{\dd\omega \, \omega^2}{{48 \pi^2}}
     \bigg[
     \abs{\tilde{h}_{2,2}^+ + \iu \tilde{h}^\times_{2,2}}^2 +  \abs{\tilde{h}_{2,2}^+ - \iu \tilde{h}^\times_{2,2}}^2  \nonumber \\
     &&&
     -\abs{\tilde{h}_{2,-2}^+ + \iu \tilde{h}^\times_{2,-2}}^2 - \abs{\tilde{h}_{2,-2}^+ - \iu \tilde{h}^\times_{2,-2}}^2 
     \bigg]
     \label{eq:vz}\raisetag{1.3\baselineskip}
\end{alignat}
{where the additional $2\pi$ factor arises as a result of performing Fourier transforms.}

In contrast to $K^J$, the leading order contribution to $\vgw$ (Eq.~\eqref{eq:VGW}) takes the following form:
\begin{alignat}{2}
    \vgw &\approx \int_0^\infty  \dd\omega  \,\omega^3 &&
     \bigg[\abs{\tilde{h}_{2,2}^+ + \iu \tilde{h}^\times_{2,2}}^2 {\color{red} -} \abs{\tilde{h}_{2,2}^+ - \iu \tilde{h}^\times_{2,2}}^2 \label{eq:vgw2} \\
     &&&{\color{red} +}\abs{\tilde{h}_{2,-2}^+ + \iu \tilde{h}^\times_{2,-2}}^2 - \abs{\tilde{h}_{2,-2}^+ - \iu \tilde{h}^\times_{2,-2}}^2 \bigg]\nonumber
\end{alignat}
Notice the relative difference of signs between the two expressions highlighted in red. Apart from sub-leading contributions not shown (which happen to be drastically different), this result shows that the two observables measure manifestly different physical quantities. When there is $m$ symmetry, in the sense of Eqs.~\eqref{eq:m1}\,\&\,\eqref{eq:m2} (\ie\ reflection symmetry in some frame), both expressions vanish, as it is the case for aligned-spin BBH systems. 
However, while $m$ asymmetry does lead to a non-zero transversal recoil velocity $K^J \neq 0$, it does not necessarily imply $\vgw\neq 0$. 

Despite the different nature of both observables, an approximate calculation reveals that, for the case of black-hole mergers, $\vgw$ and $K^J$ are closely related to each other. We will elaborate on this in App.~\ref{app:hphc_relation}, and sketch the idea here. 
By definition, $h^{+}_\lm(u)$ and $-h^{\times}_\lm(u)$ are the real and imaginary parts of a complex function $h_\lm$, which can be decomposed into some real-valued amplitude $A_{\ell,m}(u)$ and phase $\phi_\lm(u) = \omega_\lm(u) u + \phi^{0}_\lm$, as done in~\cite{NRSur4d2s}. 
Next, let us assume that $h_\lm(u)$ is an analytic function in the upper-half complex plane -- which depends on the properties of the source -- then $h^{\times}_\lm(u)$ is simply the Hilbert transform of $h^{+}_\lm(u)$~\cite{King2009:HilbertTransform,Brown1974:Analytic}. 
As shown in App.~\ref{app:hphc_relation}, this assumption is well justified for $h_{2,\pm 2}$ throughout the whole evolution of black-hole mergers\footnote{In principle, this may only be intuitive for the inspiral regime, where it is well-known that each phase of each mode depends on the orbital frequency $\omega_{\lm}$ as $m \omega_{\rm orb}$, and both amplitude and phase evolve slowly.}, excluding fine-tuned cases that display transitional precession. 
Under this hypothesis, the corresponding Fourier transforms satisfy ${\tilde h^\times}_\lm(\omega) = -i \sgn(\omega)\,{\tilde h^+}_\lm(\omega)$. Taking into account that for BBHs, the frequency $\omega_\lm$ of the positive (negative) $m$ modes is positive (negative)\footnote{Note that this is a much weaker condition ($\sgn(\omega_\lm) = \sgn(m)$) than the well-known relation $\omega_\lm \propto m \omega_{\rm orb}$ (with $\omega_{\rm orb}$ denoting the orbital frequency) which only holds during the inspiral regime.}, one obtains that {${\tilde h^\times}_\lm(\omega) =- \iu \sgn(m){\tilde h^+}_\lm(\omega)$} for $\omega > 0$. 
This above causes the terms in Eqs.~\eqref{eq:vz} \&~\eqref{eq:vgw2} that are differed by a relative minus sign to vanish, making the integrands of $K^J$ and $\vgw$ equal to each other modulo constant factors and powers of $\omega$. 

Next, note that the integral is vastly dominated by contributions from a narrow range of times around the merger time (see Fig.~\ref{fig:flux_comparison}), which correspond to a narrow range of frequencies around some value of $\omega_0$. 
With this, we can obtain a ballpark relation {(see also App.~\ref{app:derivation} for a more detailed analysis)}
\begin{equation}
    {\vgw \simeq 48 \pi^2 \, (M \omega_{0})\, K^J / c}\ , \label{eq:main_result}
\end{equation}
where {$M\omega_0 \simeq {{\cal{O}}(0.5)}$} is the characteristic dimensionless angular frequency (in geometric units), around which merger typically takes place. 
This is, $\vgw \simeq {0.790} \times K^J / (10^3 \, \si{\km\per\s}) $. 
{As we will show in Sec.~\ref{sec:testing}, this coefficient agrees with the empirical results obtained from numerical simulations. Furthermore, we have checked that the spread of $\vgw$ in Fig.~\ref{fig:SXS_regression} can be mostly attributed to the fact that the characteristic frequency depends on the particular system, see Fig.~\ref{fig:omega_distribution}, and to the omission of contributions from the interactions terms of the subdominant modes in our assumption.}

In summary, while not necessarily true for generic gravitational-wave sources, the properties of BBHs yield an approximately linear relation between $\vgw$ and $K^J$, which we will test next through numerical relativity~(NR) simulations.

\subsection{Testing correlations using numerical simulations of black-hole mergers\label{sec:testing}}
We now test the relation found in the previous section by explicitly inspecting numerical simulations of black-hole mergers. 

Although the {total angular momentum $\vb*J$} is roughly conserved throughout the evolution of the system, its value depends on the reference time at which it is defined. Due to this, we find it convenient to analyze this question by using also the projection of $\vb*{K}$ onto the spin of the final black hole, which is uniquely defined. We denote this by $K\mathscr{h}$, where $\mathscr{h}=\langle \vu*{K}, \vu*{a}_{\rm f}\rangle$ is the helicity of the final black hole\footnote{
Remarkably, most remnant BHs end up having $\mathscr{h} \approx \pm 1$ as shown in App.~\ref{app:final_spin}, which resonates with the helicity of massless particles.}. Because $\vu*{J}$ is roughly conserved in time, both $\vu*{J}$ and $\vu*{a}_{\rm f}$ are roughly parallel to each other. Therefore, both calculations are expected to yield similar results, as we will show.

In the following, we will explicitly compute $\vgw,\ K^J$ and $h$ for all of the NR simulations of black hole mergers included in both the RIT~\cite{RIT2017:1stCatalog,RIT2019:2ndCatalog,RIT2020:3rdCatalog,RIT2022:4thCatalog} and SXS catalogues~\cite{SXS2019:Catalog}, which include both precessing and eccentric black-hole mergers. We include all of the $(\ell,m)$ modes present in the simulations.

For practical purposes, it is convenient to write $\vgw$ in terms of the Newman-Penrose spin-coefficient formalism~\cite{NP62}. This is given by~\cite{dR21}
\begin{equation}
    \vgw = 2\pi\,\int_{-\infty}^{\infty} \dd u \int_{-\infty}^{u} \dd u' \sum_{\ell\,m} 
     \Im\qty[ \Psi_{4,\ell m}^0(u)\,\bar{\Psi}_{4,\ell m}^0(u')] \ ,
    \label{eq:vgw_psi4}
\end{equation}
where $\Psi_4^0=\lim_{r\to \infty} r\,\Psi_4$ is the leading order behaviour of the Newman-Penrose scalar $\Psi_4$ near future null infinity.
While conceptually less transparent than Eq.~\eqref{eq:VGW}, this expression will be more useful for evaluations with GW waveforms. %in the following sections when discussing GW waveforms.

When evaluating Eq.~\eqref{eq:vgw_psi4}, we remove the initial part of the waveforms, which is typically contaminated by an initial unphysical burst known as ``junk radiation''. 
Moreover, since the integrals that we compute are vastly dominated by the emissions near merger, this will not affect the final value of $\vgw$~\cite{dRetal20,sanchis2023precessing,toappear}.

\begin{table*}[t]
\sisetup{retain-explicit-plus}%
\centering
\resizebox{510pt}{!}{%
\begin{tabular}{|c|ccc|cccc|ccccc|c|c|}
\hline
$\ell$    & \multicolumn{3}{c|}{2} & \multicolumn{4}{c|}{3} & \multicolumn{5}{c|}{4} & \multirow{2}{*}{$\vgw$} &\multirow{2}{*}{$Q$} \\
$m$       & $\pm 2$ & $\pm 1$ & 0          & $\pm 3$ & $\pm 2$ & $\pm 1$    & 0          & $\pm 4$ & $\pm 3$ & $\pm 2$    & $\pm 1$    & 0  &  &\\ \hline
\texttt{BBH:0130}  & 90.7    & 2.0     & 0.1        & 3.6     & 1.7     & \num{1e-2} & \num{2e-3} & 1.7     & 0.3     & \num{3e-2} & \num{4e-4} & \num{4e-4} & \num{+0.866}&0.75 \\ \hline
\texttt{BBH:0168}  & 81.7    & 11.6    & 1.7        & 2.4     & 1.4     & \num{6e-2} & \num{3e-3} & 0.9     & 0.1     & \num{2e-2} & \num{3e-3} & \num{6e-4} & \num{+0.180}&0.485 \\ \hline
\texttt{BBH:0234}  & 61.9    & 21.2    & 1.6        & 9.3     & 2.5     & \num{3e-2} & $10^{-2}$  & 2.1     & 1.3     & 0.2        & $10^{-2}$  & \num{4e-4} & \num{+0.077}&0.6667 \\ \hline
\texttt{BBH:0363}  & 95.9    & 0.0     & $10^{-2}$  & 0.0     & 0.9     & 0.0        & \num{6e-4} & 4.2     & 0.0     & $10^{-2}$  & 0.0        & \num{4e-4} & \num{-3.349}&1.00 \\ \hline
\texttt{BBH:0393}  & 94.5    & 0.0     & \num{4e-4} & 0.0     & 0.6     & 0.0        & \num{5e-4} & 4.8     & 0.0     & \num{9e-3} & 0.0        & \num{4e-4} & \num{-3.659}&1.00 \\ \hline
\texttt{BBH:0504}  & 94.6    & 0.0     & \num{3e-3} & 0.0     & 0.6     & 0.0        & \num{4e-4} & 4.9     & 0.0     & \num{8e-3} & 0.0        & \num{4e-4} & \num{+3.561}&1.00 \\ \hline
\texttt{BBH:0874}  & 34.8    & 19.6    & 4.7        & 15.3    & 11.1    & 4.7        & 0.8        & 3.1     & 3.0     & 2.1        & 0.6        & \num{4e-2} & \num{+0.233}&0.06667 \\ \hline
\texttt{eBBH:1603} & 96.2    & 0.0     & \num{2e-2} & 0.0     & 0.8     & 0.0        & \num{5e-3} & 2.9     & 0.0     & \num{9e-3} & 0.0        & \num{8e-4} & \num{-2.918}&1.00 \\ \hline
\texttt{eBBH:1604} & 96.7    & 0.0     & \num{6e-2} & 0.0     & 0.6     & 0.0        & \num{3e-3} & 2.6     & 0.0     & \num{5e-3} & 0.0        & \num{5e-4} & \num{+1.520}&1.00 \\ \hline
\texttt{eBBH:1632} &  4.7    & 7.4     & 0.3        & 40.4    & 20.2    & 7.4        & 0.2        & 13.0    & 3.8     & 1.4        & 1.3        & \num{8e-4} & \num{-3.717e-6}&1.00\\ \hline
\end{tabular}
}
\caption{{\bf Relative contributions of each mode pair to $\vgw$.} 
    Each column shows the percentages of each pair of $(\lm)$ modes contributing to the value of $\vgw$ (second last column) according to Eq.~\eqref{eq:vgw_psi4} and the mass ratio $Q$. The rows correspond to ten different numerical simulations of BBH mergers from the RIT catalogue.
    The result shows that all the precessing BBHs have the quadropolar modes (2, $\pm 2$) dominate the sum in Eq.~\eqref{eq:vgw_psi4}. 
    Except for the last row, \texttt{eBBH:1632}, which corresponds to an aligned-spin binary, and the $\vgw$ value is six order-of-magnitude less the others and compatible with zero.
}
\label{tab:relative}
\end{table*}

\begin{figure*}[htb]
    \centering
    \includegraphics[width=0.48\linewidth]{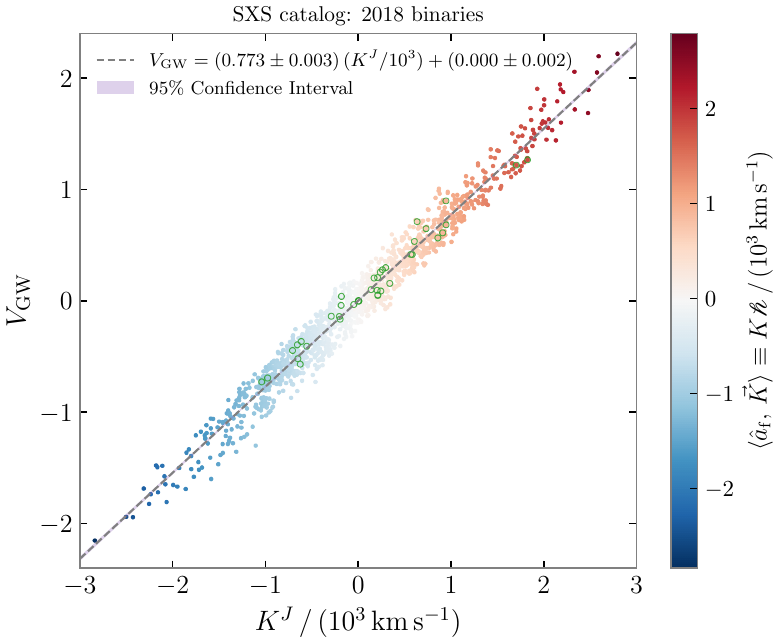}
    \includegraphics[width=0.48\linewidth]{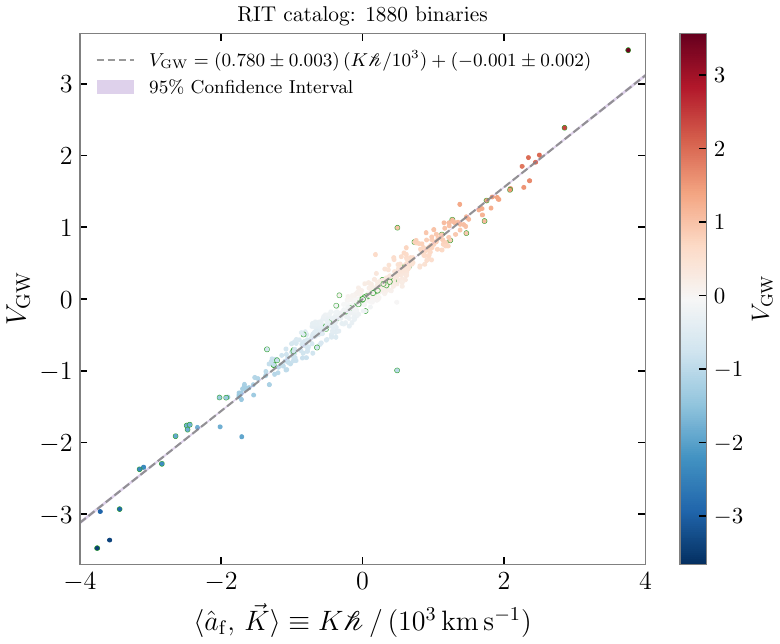}
    \caption{{\bf Strong linear correlation of $\vgw$ with $K^J$ and $K\,h$ in the SXS and RIT catalogues.}  
        In the left panel, it is showing the 2018 BBH simulations in the SXS catalogue, including eccentric and precessing binaries. For each of them, we compute the projection of the recoil velocity $\vb*{K}$ along the $\vb*J$ direction at $t = -100\,M$, against the Stokes parameter $\vgw$. 
    Their strong linear correlation is shown by the grey dashed line and the narrow violet shaded area, which denote the best-fit line and its 95\% confidence interval. 
        The green circles highlight the eccentric binaries, which have eccentricity greater than 0.001 at the reference time of the simulation.
    To echo Fig.~\ref{fig:linear_relation}, we colour each point by the rescaled helicity $K\,\mathscr{h}$ of the remnant BH.
    In the right panel, we plot the helicity $K\,\mathscr{h}$ against  $\vgw$ for the 1880 binaries in the full RIT catalogue, in which 824 of them are eccentric binaries which are circled in green.
    From both catalogues, it is evident that both of them share a similar slope, and their $y$-intercepts are compatible with 0.000.
     }
    \label{fig:SXS_regression}
\end{figure*}

Table~\ref{tab:relative} shows the relative contribution of each $(\ell, m)$ mode to the $V$-Stokes parameter for a sample of BBHs from the RIT catalogue. The result shows that, for all precessing binaries, the quadrupolar modes ($\ell=2,\,m=\pm2$) dominate the sum in equation~\eqref{eq:vgw_psi4}. Thus, the actual value of $\vgw$ is not significantly affected by using a sufficiently high cutoff for $\ell$\footnote{The quadrupolar modes do not necessarily dominate the sum in Eq.~\eqref{eq:vgw_psi4} for non-precessing BBHs, as shown by the row of \texttt{eBBH:1632} in Table~\ref{tab:relative}. However, these are aligned-spin systems for which $\vgw$ happens to be highly suppressed (with values compatible with zero within numerical error) due to cancellation among positive and negative $m$ modes. }. 

The left panel of Fig~\ref{fig:SXS_regression} shows the value of $\vgw$ as a function $K^J$ for the 2018 BBHs present in the \texttt{SXS} catalogue. The unit vector $\vu*{J}$ is computed at a time $t=-100\,M$ before the merger. A clear linear correlation can be observed, consistent with the expectation discussed in Sec.~\ref{sec:theory_expectation}. The color denotes the value of the helicity, which also follows a linear relation with both $\vgw$ and $K^J$. Such a relation is made more clear in Fig.~\ref{fig:linear_relation}, which shows the helicity as a function of $K^J$, with the color denoting the value of $\vgw$. In this plot, {it is clear that the scaled helicity ($K\mathscr{h}$) has an almost one-to-one correspondence, reflecting the intuitive result that the final spin tends to be aligned with the total angular momentum, which is expected except for very fine-tuned cases which undergo transitional precession. We will further explore this alignment  in App.~\ref{app:final_spin}. Furthermore, the color distribution shows that the sign of the helicity (the projection of the final recoil onto the final spin) determines the sign of $\vgw$, which resonates with the results in Fig.~\ref{fig:SXS_regression}.}

For completeness, we also computed $\vgw$ for all simulations in the RIT catalogue and plotted it against the helicity on the right of Fig.~\ref{fig:SXS_regression}, coloured by the value of $\vgw$. The same linear relationship is found, also with a $y$-intercept at nearly zero\footnote{The fact that both $V_{\rm GW}$ and $\mathscr{h}$ show equal signs is somewhat consistent with previous work~\cite{Zhu_Harrison_qnms_precession} showing that $\mathscr{h}$ is related to the ratio of the $(2,2)$ and $(2,-2)$ modes, although such study was restricted to the ringdown stage of quasi-circular, nearly equal-mass binaries.}. We remark that unlike the \texttt{SXS} catalogue, neither $\vb*J$ nor $\vb*a_{\rm f}$ can be computed from the public data available in the \texttt{RIT} catalog directly. In order to compute $\< \vu*a_{\rm f},\,\vb*K>$ for the $x$-axis, we estimated $\vb*a_{\rm f}$ by subtracting the net change of angular momentum from the initial total angular momentum.

\begin{figure}
    \centering
    \includegraphics[width=1\linewidth]{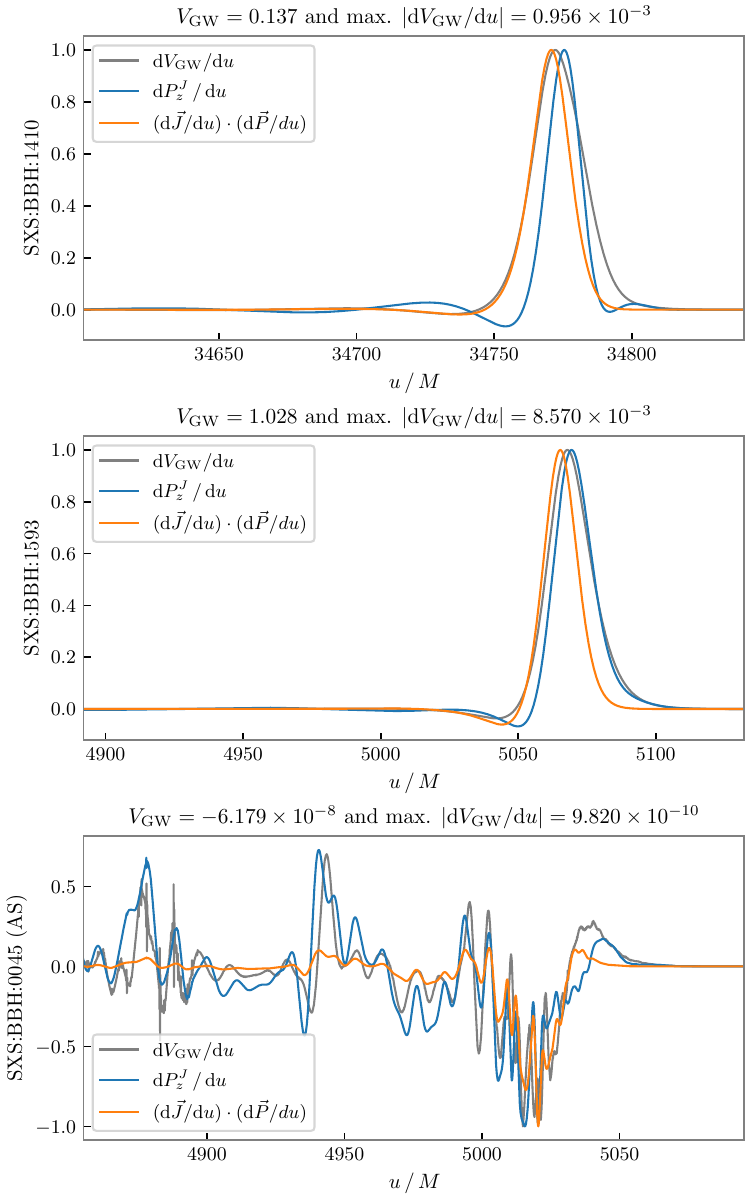}
    \caption{{\bf The high similarities of the three types of fluxes.} 
    The three panels from top to bottom show the fluxes integrated from three selected \texttt{SXS} simulations. 
    The first two correspond to precessing binaries, whereas the last is an aligned-spin binary.
    The curve in grey is $\dd \vgw\,/\,\dd u$, that is the integrand of Eq.~\eqref{eq:vgw_psi4}. 
    The blue curve is the component of the linear momentum flux in the $\vu*J$ direction, computed at $t_{\rm ref} = -100\,M$.
    Finally, the orange curve is the inner product between the angular and linear momenta fluxes.
    For easy comparison, all three fluxes are normalised with respect to their absolute peak value.
    Note that in these three panels, the bulks of emission are all roughly located within a $50\,M$ window near merger.
    }
    \label{fig:flux_comparison}
\end{figure}

The above discussion pertains to ``fixed'' or ``integrated'' quantities. To provide further detail on the evolution of such quantities throughout the evolution of the binary, Fig.~\ref{fig:flux_comparison} shows the values of $\dv*{\vgw}{u}$, $\dv*{K^J}{u} $ as a function of time for three different cases. The top two panels correspond to precessing BBHs, while the bottom one corresponds to an aligned-spin case. Again, a clear correlation can be observed between the two quantities, modulo certain time delays. 

Due to the chiral nature of $\vgw$ and its relation with the helicity of the remnant BH, one may expect it may correlate with how helicity is emitted from the system. Motivated by this, we define the ``helicity flux'' as $\< \dv*{\vb*J}{u},\,\dv*{\vb*K}{u} >$, and is plotted in orange. 
We find that its emission also correlates with that of $\vgw$, and peaks at nearly the same time, confirming the expectation.

\section{Discussion}
\label{sec:Discussion}

The spin configurations of the BBHs can determine whether a spacetime is mirror symmetric at an instance. 
In particular, those with parallel spins (anti-)aligned with orbital angular momentum are mirror symmetric at all times, therefore leading to a vanishing $\vgw$.
On the other hand, $\vgw \neq 0$ necessarily implies a mirror asymmetric spacetime. 
That said, BBHs with parallel spins but misaligned with orbital momentum will precess and display a small but non-trivial $\vgw$, even though they can be mirror symmetric momentarily.
Interestingly, these mirror asymmetric spacetimes will leave an imprint on the GW they emit, precisely in terms of the imbalance between right- and left-handed circularly polarised GW fluxes, which is what $\vgw$ measures.
In this work, we further extend this relation and relate it with the helicity of the remnant black hole, another chiral quantity.
In other words, we have explicitly connected the mirror asymmetry in the spin configurations, with the imbalance of circularly polarised flux and finally with the helicity of the final black hole.
Intuitively, these three are understood to be deeply intertwined by the fact that they are different perspectives on the intrinsic chirality of the BBHs\footnote{{The spin vectors used in this study are obtained from the horizon data that are released with the numerical simulations in the catalogues. Although there is no manifestly invariant definition of spin vectors~\cite{Szabados2009:quasilocal} and hence possible gauge dependences, given the linear correlation found with $\vgw$, which is a gauge-invariant quantity, it is reasonable to expect that the inner product of these vectors is physically meaningful.}
}.

The study of the net flux of circularly polarized waves across the whole population of black-hole mergers can reveal hidden, non-manifest, asymmetries in our Universe. For instance, if the Cosmological Principle holds, then the average flux $\<\vgw>$ of BBH population should be consistent with zero, even if individual sources yield non-zero $\vgw$. Such study was indeed carried out in Ref.~\cite{toappear} over 47 BBH detections from the third observing catalogs of Adv.~LIGO and Adv.~Virgo, yielding consistency with the Cosmological Principle, albeit with large uncertainties due to the limited statistics.
Moreover, due to its relation to orbital precession, the study of mirror asymmetry for BBHs in our Universe can provide valuable information about BH astrophysical formation channels.

Further and louder observations of BBHs will enable more accurate tests of the cosmological principle through measurements of $\vgw$. In anticipation of this, we have shown through the recovery of numerically simulated signals that accurate measurements of $\vgw$ can be performed using the waveform model, \nrsur, in the near-mid future even for BBH systems with SNRs of 50, way beyond current detections. This further confirms the robustness of the results presented in~\cite{toappear}.

Next, for the case of binary black-hole mergers, we have shown that, despite its highly non-trivial expression as an integrated quantity through the evolution of the BBH, $\vgw$, is linearly correlated to the helicity of the final black hole, understood as the projection of its final recoil onto its final spin, as shown in Figs.~\ref{fig:linear_relation}~\&~\ref{fig:SXS_regression}. Such linear relation is {\it a priori} far from obvious, as both quantities are obtained through manifestly different expressions. For the case of BBHs, we have first derived an approximate relation analytically -- under reasonable assumptions -- to then confirm it numerically by explicitly computing both quantities for a wide range of numerical simulations. 

The discussed relation suggests an interesting conceptual parallelism with the experiment that lead to the discovery of parity broken by the weak force. According to our findings, the experiment of measuring $\vgw$ across a large sample of BBH observations may be thought of as a cosmological analogue of Wu's experiment in particle physics~\cite{PhysRev.105.1413}. This experiment targeted the parity symmetry breaking of the weak interaction, which governs the spontaneous decay of particles. By analyzing the beta decay of an ensemble of Cobalt atoms, and in particular the linear momentum distribution with respect to the spin of the host atoms (which provides a notion for the helicity of the emitted electrons), a preferred direction of emission was found, thereby indicating a failure of mirror symmetry in weak interactions. 
Here, the spin of the final BH would play the same role as the spin of the Cobalt atoms, while the recoil of the remnant BH plays the role of the electron linear momentum. Therefore, although BBHs cannot be regarded as fundamental quantum systems, remnant BHs can be similarly used to study the mirror symmetry of the whole Universe, when understood as an ensemble of all BBH sources. By measuring the distribution of the linear momenta of remnant black holes with respect to their own spin, which is a notion for the helicity of the ``emitted'' BH after the merger, we have the potential to reveal underlying global asymmetries in our Universe. 
This concept was similarly applied to the distributions of spiral galaxies in Refs.~\cite{Longo:2007gr,Longo:2011nk}.

A first study of this idea on BBH was carried out using $\vgw$ in~\cite{toappear}, which, as we show here, is equivalent to the helicity of the binary. 
Notably, such potential averaged asymmetry may be sourced by either a fundamental property of gravitational interactions beyond GR, or simply because BBH formation scenarios favour a given helicity sign. 
In~\cite{toappear} it was assumed that the observed data fit the waveforms as predicted by standard General Relativity, but the method can be equivalently applied to the predictions of any other modified theory of gravity, \eg\ dynamical Chern-Simons gravity, which is a fundamentally chiral theory.  Forthcoming BBH observations will enable us to determine whether our Universe has a preferred handedness.

\begin{acknowledgments}

We are grateful to Koustav~Chandra and Maria~Haney for their review and comments on the manuscript. 
SHWL acknowledges support by grants from the Research Grants Council of Hong Kong (Project No.~CUHK~14304622 and 14307923), and the Direct Grant for Research from the Research Committee of The Chinese University of Hong Kong. 
AFT acknowledges support from ``Ajudes d'iniciaci\'o a la investigaci\'o'' funded by the {\it Departament d'Astronomia i Astrof\'isica}, Universitat de Valencia. NSG and JCB respectively acknowledge support from the Spanish Ministry of Science and Innovation via the Ram\'on y Cajal programme (grants RYC2022-037424-I and RYC2022-036203-I), funded by MCIN/AEI/10.13039/501100011033 and by ``ESF Investing in your future”. 
JCB is also supported by the research grants PID2020-118635GB-I00 and PID2024-160643NB-I00 from the Spain-Ministerio de Ciencia e Innovaci\'{o}n and the grant ED431F 2025/04 of the Galician CONSELLERIA DE EDUCACION, CIENCIA, UNIVERSIDADES E FORMACION PROFESIONAL. JCB also acknowledges support by the programme HORIZON-MSCA2021-SE-01 Grant No. NewFunFiCO-101086251. IGFAE is supported by the Ayuda Maria de Maeztu CEX2023-001318-M funded by MICIU/AEI /10.13039/501100011033.
NSG is further supported by the Spanish Agencia Estatal de Investigaci\'on (grant PID2021-125485NB-C21) funded by MCIN/AEI/10.13039/501100011033 and ERDF A way of making Europe. We acknowledge further support by the European Horizon Europe staff exchange (SE) programme HORIZON-MSCA2021-SE-01 Grant No. NewFunFiCO-101086251.
We acknowledge the use of computing facilities supported by grants from the Croucher Innovation Award from the Croucher Foundation Hong Kong.
ADR acknowledges financial support via  ``{\it Atraccion de Talento Cesar Nombela}'', grant No 2023-T1/TEC-29023, funded by Comunidad de Madrid, as well as support from the project PID2023-149560NB-C21 funded by MCIU /AEI/10.13039/501100011033 / FEDER, UE.
This manuscript has LIGO-DCC number P2500013.

\end{acknowledgments}

\appendix

\begin{appendix}

\section{Further justification of the usage of $M\omega = 0.5$ in Equation (12) \label{app:derivation}}
In the main text, we argue that Eq.~\eqref{eq:main_result} is valid due to the fact that the integrands in equations~\eqref{eq:vz} and~\eqref{eq:vgw2} are highly peaked at a particular value of the frequency $\omega$ typically given by $M\omega_0 \simeq 0.5$.
In this appendix, we provide a more detailed analysis of the equations for $\vgw$ and $K^J$ that will lead to the same quantitative result through several approaches.

By comparing the like-terms in the integrands of Eqs.~\eqref{eq:vz} \&~\eqref{eq:vgw2}, one arrives at the following equation:
\begin{equation}
    \dv{\vgw}{\omega} = 48\pi^2 \,\omega\dv{K^J}{\omega} \ ,
\end{equation}
where the two derivatives w.r.t.\,$\omega$ correspond to the integrands in the respective equations. Also, recall that the discussion here only pertains to the dominant $(2,\ \pm 2)$ modes. 

Integrating both sides yields:
\begin{align}
    \vgw &= \int_0^\infty \dv{\vgw}{\omega}\,\dd\omega \ , \\
    &= 48 \pi^2 \int_0^\infty \omega \dv{K^J}{\omega}\,\dd\omega\ , \\
    &= 48 \pi^2 \lim_{\omega' \to \infty} \qty{\omega K^J(\omega) \eval_0^{\omega'} - \int_0^{\omega'} K^J(\omega)\,\dd\omega }\ .~\label{eq:IbP}
\end{align}
Note that $K^J(\omega)$ is defined as the integral of $\dv*{K^J}{\omega}$ up to a maximum, finite value of $\omega$.
This is motivated by the fact that, after a certain cut-off value $\omega_0$, the final remnant BH settles to its final state and ceases to emit GWs. This makes $\dv*{K^J}{\omega}$ vanishes causing $K^J(\omega)$ to become a constant which we denote by $K^J_0 = K^J(\omega_0)$.

\begin{figure}[tpb]
    \centering
    \includegraphics[width=1\linewidth]{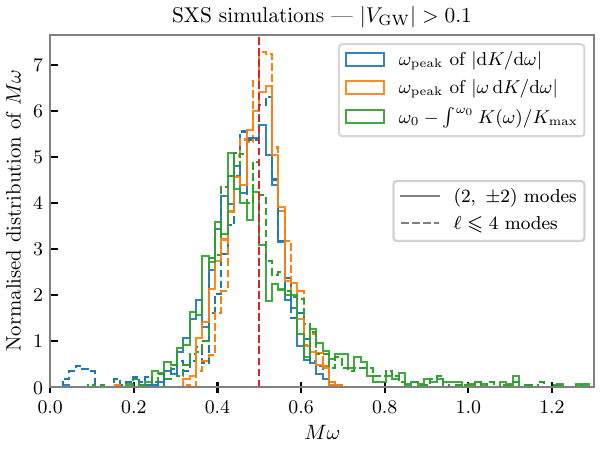}
    \caption{
    {\bf Distributions of the dimensionless characteristic frequencies for all \texttt{SXS} BBH simulations with $\abs{\vgw} > 0.1$.} 
    The three different colours denote the different proxies for the characteristic frequencies in Eq.~\eqref{eq:main_result}. Curves in blue and orange are the frequencies at which $\abs{\dv*{K^J}{\omega}}$ and $\abs{\omega\,\dv*{K^J}{\omega}}$ attain their maxima; while the green curve denotes the bracket term in Eq.~\eqref{eq:proper_linear_relation}. Computations of $K(\omega)$ including only the dominant $(2,\pm2)$ modes and all $\ell \leqslant 4$ modes are shown with solid and dashed curves respectively.
    We show that in all cases, these frequencies are clustering around 0.5.
    }
    \label{fig:omega_distribution}
\end{figure}

With the above, the two boundary terms can be further simplified. The one at $\omega'$ becomes $\omega' K^J_0$ and the other vanishes at zero. Furthermore, the second integral can be expanded into:
\begin{equation}
    \int_0^{\omega'} K^J(\omega)\,\dd\omega = \int_0^{\omega_0} K^J(\omega)\,\dd\omega + (\omega' - \omega_0)\,K^J_0\ .
\end{equation}
Combining all these, the diverging term, $\omega' K_0^J$, at the boundary cancels with the upper limit of the integral in Eq.~\eqref{eq:IbP}, and we arrive at:
\begin{align}
    \vgw &= 48\pi^2 \qty[
            \omega_0 K^J_0 - \int_0^{\omega_0} K^J(\omega)\,\dd\omega
        ] \ ,\\ 
    &= 48\pi^2 \qty[
    \omega_0 - \int_0^{\omega_0} \frac{K^J(\omega)}{K_0^J}\,\dd\omega
    ] K^J_0 \ . 
    \label{eq:proper_linear_relation}
\end{align}
Importantly, note that the variable $\omega_0$ does not have the same definition as in the main text, where it denoted the peak emission frequency. Nevertheless, we found empirically that the bracket in Eq.~\eqref{eq:proper_linear_relation} also takes value around ${\cal O}(0.5)$. This was found by computing the value of this bracket for all the \texttt{SXS} simulations used in the main text.  Fig.~\ref{fig:omega_distribution} shows the corresponding result, clearly showing this has as a similar distribution as the peak emission frequency. This further strengthens the validity of the approximate relation, $\vgw \approx 24 \pi^2 K^J$, and explains partly the spread seen in the linear relation in Fig.~\ref{fig:SXS_regression}. 
In Fig.~\ref{fig:omega_distribution}, we compare the peak emission frequencies with the expression in bracket. Note that we show only results for cases with $\abs{\vgw} > 0.1$. The reason is that the net emissions of $K^J(\omega)$ are otherwise vanishingly small for near aligned-spin systems, so that one cannot reliably define a notion of ``emission peak'' for our integrals. We find that although our analytic results are derived by considering only the dominant quadrupolar modes, the subdominant contributions from the higher order modes do not affect this result qualitatively. This is reflected by the overlapping distributions displayed with solid and dashed curves, which respectively correspond to the results computed with only the dominant quadrupolar modes and with all $\ell \leqslant 4$ modes, respectively.

Finally, to completely reconcile with the empirical results, recall that the conversion between the geometric and SI units of the kick is $K^J = K_{\rm SI} / c$.

\section{The role of asymmetric modes in $\vgw$ \label{app:asym_modes}}
Parity asymmetry is most often tied with the $m$-asymmetric modes in most literatures~\cite{Boyle2014:Precession,Mielke:2024kya}.
In view of this, it is worth exploring the relation between $\vgw$ and the conventional asymmetric modes. 
In this section, we will express $\vgw$ in terms of the asymmetric modes, which will also lead to an expression that closely resembles that of the recoil. 
Following Refs.~\cite{Mielke:2024kya}, we define the symmetric and asymmetric modes as:
\begin{equation}
    h^\pm_\lm(u) = \frac{1}{2}\qty[ h_\lm(u) \pm (-1)^\ell\,\overline{h_\lmm}(u) ]
\end{equation}
\textit{Note that throughout this section, and this section only}, we will use $h^+_\lm$ to denote the symmetric modes, and not the real part of $h_\lm$ as in the main text.
With these definitions, one can relate the conjugate of $h_\lm$ with $h_\lmm$ as:
\begin{equation}
    \overline{h_\lm} = (-1)^\ell \qty[ h^+_\lmm - h^-_\lmm]\ .
\end{equation}
\begin{widetext}
Then, denoting by $ {\cal F}$ the Fourier operator, we can rewrite the definition of $\vgw$ (Eq.~\eqref{eq:VGW}) in terms of $h^+$ and $h^-$:
\begin{align*}
    \vgw &= \int_0^\infty  \,\dd  \omega \,\omega^3 \sum_{\ell = 2}^\infty \sum_{m = -\ell}^\ell\qty [
    \abs{ \vphantom{\overline{h}} {\cal F}\qty[\overline{h_\lm}](\omega)}^2 - \abs{ {\cal F}\qty[h_\lm](\omega)}^2  ] \\
    &= \int_0^\infty  \,\dd  \omega \,\omega^3 \sum_{\ell = 2}^\infty \sum_{m = -\ell}^\ell\qty [
    \abs{ {\cal F}\qty[(-1)^\ell \qty(h^+_\lmm - h^-_\lmm)](\omega)}^2 - \abs{ {\cal F}\qty[h^+_\lm + h^-_\lm](\omega)}^2  ] \\
    \intertext{Expanding both terms, replace the Fourier operator with the overhead tilde, and dropping the explicit reference to $\omega$, we arrive at:}
    &=  \int_0^\infty  \,\dd  \omega \,\omega^3 \sum_{\ell = 2}^\infty \sum_{m = -\ell}^\ell 
        \qty{
        \abs{\widetilde{h^+_\lmm}}^2 + \abs{\widetilde{h^-_\lmm}}^2 - 2\,\Re\qty[ \widetilde{h^+_\lmm} \overline{\widetilde{h^-_\lmm} }]
        } - \qty {
        \abs{\widetilde{h^+_\lm}}^2 + \abs{\widetilde{h^-_\lm}}^2 + 2\,\Re\qty[ \widetilde{h^+_\lm} \overline{\widetilde{h^-_\lm} }] 
            } \\
    &=  \int_0^\infty  \,\dd  \omega \,\omega^3 \sum_{\ell = 2}^\infty \sum_{m = -\ell}^\ell 
        \qty( \abs{\widetilde{h^+_\lmm}}^2 - \abs{\widetilde{h^+_\lm}}^2 )
      + \qty( \abs{\widetilde{h^-_\lmm}}^2 - \abs{\widetilde{h^-_\lm}}^2 )
      - 2\,\Re\qty[ \widetilde{h^+_\lmm} \overline{\widetilde{h^-_\lmm}} + \widetilde{h^+_\lm} \overline{\widetilde{h^-_\lm} } ]
\end{align*}
The first two parentheses are cancelled by summation over $m$; simplifying the expression, we obtain
\begin{equation}
    \vgw = -4\,\int_0^\infty  \,\dd  \omega \,\omega^3 \sum_{\ell = 2}^\infty \sum_{m = -\ell}^\ell  
    \Re\qty[ \widetilde{h^+_\lm} \overline{\widetilde{h^-_\lm} }]\ .
    \label{eq:vgw_asym}
\end{equation}
Note the striking resemblance to the formula of the linear momentum (Eq.~(14) in Ref.~\cite{Mielke:2024kya}). Both expressions show explicitly that the source of $\vgw$ and $\vb*K$ is the interaction between the symmetric and asymmetric modes; or equivalently, the interaction between parity preserving and violating modes~\cite{Boyle2014:Precession}. 
The main difference between the two expressions is that the expression of the recoil involves an additional integration along a particular direction, and that the expression for $\vgw$ is in the Fourier domain. 
\section{The approximate relation between $h^+_\lm$ and $h^\times_\lm$: validity and implications}
\label{app:hphc_relation}

In this appendix, we elaborate on how the assumptions we made on $h^+_\lm$ and $h^\times_\lm$ in the main text lead to our conclusions. Next, we will show both through analytic and empirical results that such assumptions are safe for the case of BBHs. We recall that we define $h^+_\lm = \Re[ h_\lm]$ and $h^\times_\lm = - \Im[h_\lm]$. We will show that for $m \neq 0$, the following relation approximately holds
\begin{equation}
    h^\times_\lm(u) \approx \sgn(m)\,H[h^+_\lm](u) \qqtext{and} h^+_\lm(u) \approx -\sgn(m)\,H[h^\times_\lm](u)\ .
\end{equation}
Above, $H[\cdot]$ represents the Hilbert transform, and the second relation follows directly from the inversion property of the Hilbert transform~\cite{[{}][{, most of the quoted results can be found in Chapter 4 therein.}]King2009:HilbertTransform}. The above relation implies that the respective Fourier transforms are related by:
\begin{equation}
    \widetilde{h^\times_\lm}(\omega) \approx -\iu\sgn(m\omega)\,\widetilde{h^+_\lm}(\omega) 
    \qqtext{and}
    \widetilde{h^+_\lm}(\omega) \approx +\iu\sgn(m\omega)\,\widetilde{h^\times_\lm}(\omega) \ .
    \label{eq:HilbertFourier}
\end{equation}
Notice that the terms in the integrands of Eqs.~\eqref{eq:vz} \& \eqref{eq:vgw2} will become:
\begin{align}
     \abs{\widetilde{h^+_\lm}(\omega) + \iu \widetilde{h^\times_\lm}(\omega) }^2  
     &\approx \abs{\qty(1 + \sgn(m\omega))\,\widetilde{h^+_\lm}(\omega)}^2 \\
     \abs{\widetilde{h^+_\lm}(\omega) - \iu \widetilde{h^\times_\lm}(\omega) }^2  
     &\approx \abs{\qty(1 - \sgn(m\omega))\,\widetilde{h^+_\lm}(\omega)}^2 
\end{align}
given that only positive $\omega$ will be involved in the integral, each $\sgn(m\omega)$ reduces to $\sgn(m)$.

In Eqs.~\eqref{eq:vz} \& \eqref{eq:vgw2}, under this assumption, the two terms with alternative signs will vanish.
In other words, the contribution from each $(\ell, \pm m)$ pair in both integrands can be reduced to the same expression:
\begin{equation}
    \abs{\widetilde{h^+_\lm}(\omega) + \iu \, \widetilde{h^\times_\lm}(\omega)}^2  
    - \abs{\widetilde{h^+_\lmm}(\omega) - \iu \, \widetilde{h^\times_\lmm}(\omega)}^2  
\end{equation}

\end{widetext}
\subsection{Validity of the approximate relation}
\begin{figure}[ht!]
    \centering
    \includegraphics[width=0.48\textwidth]{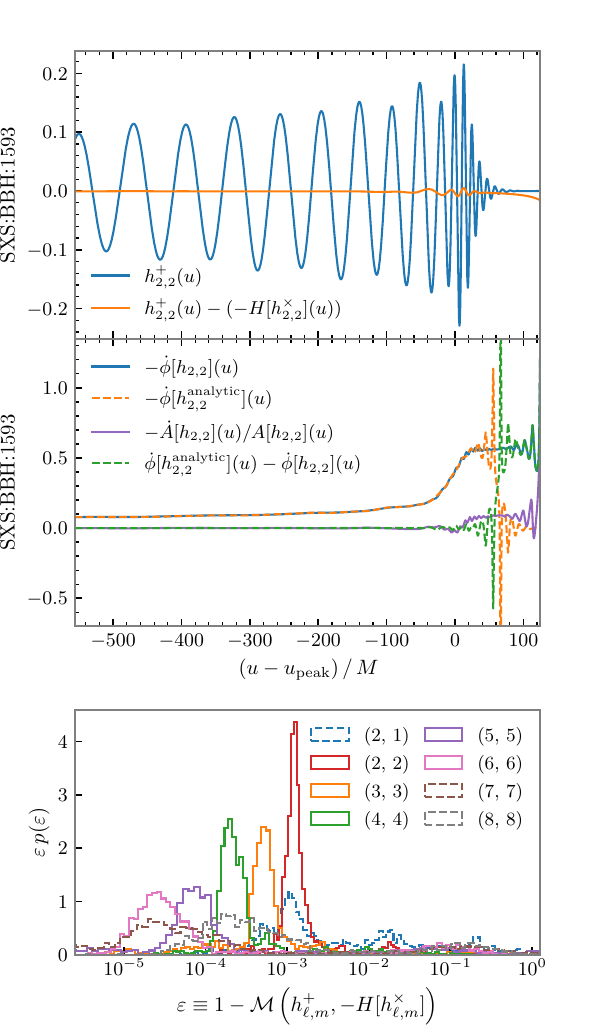}
    \caption{{\bf Validity of the Hilbert transformed approximation.} 
        From top to bottom, the first plot shows the difference between $h^+_{2,2}$ and the (negative of) Hilbert transform of $h^\times_{2,2}$, which should approximate $h^+_{2,2}$ itself under the assumption, as a function of time $u$. 
        The second panel shows the angular frequencies of $h_{2,2}$ and the analytic signal constructed from $h^+_{2,2}$, \ie~$h_{2,2}^{\rm analytic} = h^+_{2,2} - \iu H[h^+_{2,2}]$, in blue and orange respectively. Their difference is denoted by the green dashed line. The purple curve shows the derivative of the amplitude of $h_{2,2}$, \ie~$\dot A[h_{2,2}]$, scaled by the amplitude; where it deviates from zero is correlated to where the difference in $\dot \phi$ is large.  
        In the bottom panel, it shows the distribution of the time domain mismatch, $\varepsilon$, between $h^+_{\ell, m}$ and its Hilbert approximation for all $\ell = m$ modes beginning from $\ell = 2$ to 8, plus a (2, 1) modes, of all waveforms in the \texttt{SXS} catalogue, in red, orange, green and blue respectively. The typical mismatch is roughly below \num{2e-3}, except for the (2, 1) mode.
    }
    \label{fig:verify_hilbert_approx}
\end{figure}

To show that our assumption $h^\times_\lm \approx \sgn(m)\,H[h^+_\lm]$ is well-justified, we proceed by first assuming each $h_\lm$ can be described as an oscillatory signal with a slowly-varying amplitude, \ie
 \begin{equation}
     h_\lm(u) \approx A_\lm(u)\,\exp[-\iu \, \omega_\lm u]\ ,
\end{equation}
where $\omega_\lm$ is the GW angular frequency of each  $ (\lm)$ mode, with the condition $\dot{A}_\lm / A_\lm \ll \omega_\lm$.
Then by definition, its real and imaginary components are given by:
\begin{align}
    h^+_\lm &= A_\lm(u)\,\cos(\omega_m u) \\
    h^\times_\lm &= A_\lm(u)\,\sin(\omega_m u ) \ .
\end{align}
Under the slow-varying amplitude conditions (see Fig.~\ref{fig:verify_hilbert_approx}), Bedrosian's theorem~\cite{Bedrosian1963} ensures that: 
\begin{align}
    H[h^+_\lm](u) &\approx A_\lm(u) H[\cos(\omega_m u )] \\
               &\approx A_\lm(u) \sgn(m)\sin(\omega_m u ) \\
               &= \sgn(m)\,h^\times_\lm(u)\ .
\end{align}
The second line follows from the fact that Hilbert transform depends on the sign of $\omega_m$. Then taking into account that $\omega > 0$, we have $\sgn(\omega_m) = \sgn(m)$. 
This concludes $h^\times_\lm = \sgn(m)\,H[h^+_\lm]$, and similarly for $h^+_\lm$.
Moreover, as we have shown before, to explain most of the linear relationship between $\vgw$ and  $K^J_z$, we only need this relation to hold strongly for the dominant modes, which is indeed the case as we show in Fig.~\ref{fig:verify_hilbert_approx}.  

Our assumption that $h^\times_\lm \approx \sgn(m)\,H[h^+_\lm]$ implies that the complex mode $h_\lm$ satisfies~\cite{King2009:HilbertTransform,Brown1974:Analytic}:
\begin{equation}
    h_\lm \approx h^+_\lm -\iu\,\sgn(m)H\qty[h^+_\lm]\ .
    \label{eq:hlm_approx_analytic}
\end{equation}
and the following property of an analytic signal would be true:
\begin{equation}
    \widetilde{H[h_\lm]} = \iu\,\sgn(m\omega) \, \widetilde{h_\lm}\ ,
    \label{eq:FourierHilbert_hlm}
\end{equation}
where the overhead tilde denotes Fourier transformation.

In the remainder of this section, we show that this relation holds to a very high degree for precessing BBH systems in two ways. While we have performed these tests for a large suite of both numerical simulations and waveforms generated with the \nrsur\ waveform model, we highlight here results obtained on the precessing simulation \texttt{SXS:BBH:1593}.

\begin{figure}[ht]
    \centering
    \includegraphics[width=0.48\textwidth]{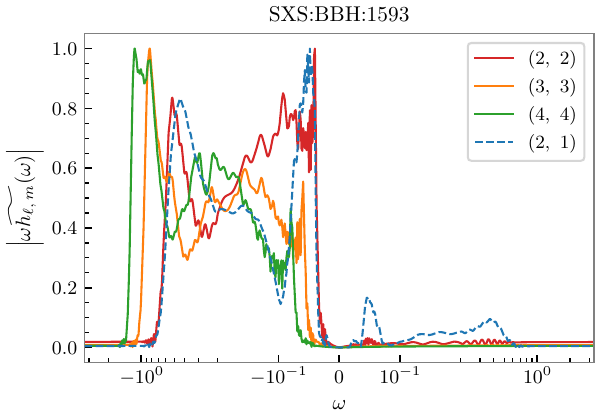}
    \caption{{\bf The vanishing positive frequency spectrum of $h_\lm$.} 
        The amplitude of the Fourier transform of $h_\lm$ for the (2, 2), (3, 3), (4, 4) and (2, 1) modes are plotted against the dimensionless frequency. Note that the spectrum is scaled by the frequency to highlight regions that are important under integration. All four curves are rescaled by their peak value. It is worth noting that other than $\tilde h_{2, 1}$, the rest have almost trivial positive frequency spectra, which implies that they behave much like analytic signals.
    }
    \label{fig:analytic_hlm_FD}
\end{figure}

First, in Fig.~\ref{fig:analytic_hlm_FD}, we show that the positive frequency spectrum of a numerically simulated complex $h_\lm$ vanishes, as expected for an analytic signal, except for the case of the $(2,1)$ mode, which we comment in the next paragraph. 

Next, Fig.~\ref{fig:verify_hilbert_approx} shows a family of tests involving the original $h_\lm$ mode and the corresponding analytic signal, constructed as $h^{\rm analytic}_\lm := h^+_\lm - \iu\,\sgn(m) H[h^+_\lm]$ according to Eq.~\eqref{eq:hlm_approx_analytic}. First, the top panel shows the original $h_{2,2}^+$ together with its minimal differences with respect to the Hilbert transform of its imaginary part $-H[h^\times_\lm]$. The green curve in the middle panel shows the minimal differences between the frequencies of the original $h_\lm$ mode $\dot{\phi}_{2,2}$ (blue) and that of the ``analytic'' mode $\dot{\phi}^{\rm analytic}_{2,2}$ (orange). 
In addition, the purple curve shows that the slowly varying condition for the amplitude is well satisfied until well into the ringdown, when the amplitude becomes exponentially decaying. As we will show later, the fact that the $(2,1)$ mode has roughly half the frequency of the $(2,2)$ makes this deviation more prominent, leading to the non-null positive-frequency components shown in Fig.~\ref{fig:analytic_hlm_FD}.
We note that in the figure, $u$ is the retarded time and $u_{\rm peak}$ is the peak time at which the norm of the waveform, $\sum_{\lm} \abs{h_\lm}^2$, is maximised.

Last, the bottom panel shows the distribution of the mismatch of $h_\lm^+$ and its respective $h_\lm^{+,\,{\rm analytic}}$ for all simulations available in the public \texttt{SXS} catalogue. 
The mismatch is defined as $\varepsilon = 1 - {\cal M}$, where $ {\cal M}$ denotes the match between two waveforms and is defined as:
\begin{equation}
    {\cal M}(a,\,b) = \frac{{\cal O}(a, b)}{\sqrt{{\cal O}(a, a)\,{\cal O}(b, b)}}\ ,
\end{equation}
and $\cal O$ is the overlap:
\begin{equation}
    {\cal O}(a, b) = \int_{-\infty}^{\infty} a(t)\,b(t) \,\dd t
\end{equation}
for real-valued functions $a(t)$ and $b(t)$.
The panel shows that for most binaries, the Hilbert approximation is valid for the dominant $ (2,2)$ mode, as well as other $m = \ell$ modes, and their mismatch is typically around or below  $10^{-3}$. 
In particular, as $\ell$ increases, the match tends to become better until $\ell = 7\;{\rm or}\;8$, where the waveform is dominated by noise.
This phenomenon can be explained as follows: the requirement for the slowly varying amplitude is equivalent to having the spectrum of the amplitude itself be disjoint from that of the oscillation, which is characterised by $\omega_m$. Higher values of $m$ imply that the two spectra are further apart, making the assumption more valid. 
On the other hand, as $m$ decreases, the spectra are closer and eventually overlap, especially when it is down to $m = 1$, which also explains the dashed-blue curve in Fig.~\ref{fig:verify_hilbert_approx}. From the physical point of view, the above is simply caused by the fact that while the amplitude of all modes varies at a similar pace, the frequency $w_{\ell,m}$ is roughly proportional to $m$, making the slowly-varying amplitude condition less (more) robustly satisfied for low (high) values of $m$.

We remark that although this approximation almost always holds for the (2, $\pm$2) mode, we have found some corner cases with a relatively high mismatch ($\varepsilon > 0.1$). These usually correspond to systems undergoing transitional precession near merger, making $\vb*J$ and $\vb*L$ nearly orthogonal, which leads to a fast amplitude change that violates the slowly varying condition. 

\section{Impact of eccentricity and other parameters}
\label{sec:eccentricity}

It is well known that the addition of orbital eccentricity to precessing systems can lead to increased kick magnitudes. In this appendix, we explore how eccentricity and other astrophysically relevant parameters like peak luminosity and final spin are related to $\vgw$.

\begin{figure}[htpb]
    \centering
    \includegraphics[width=0.49\textwidth]{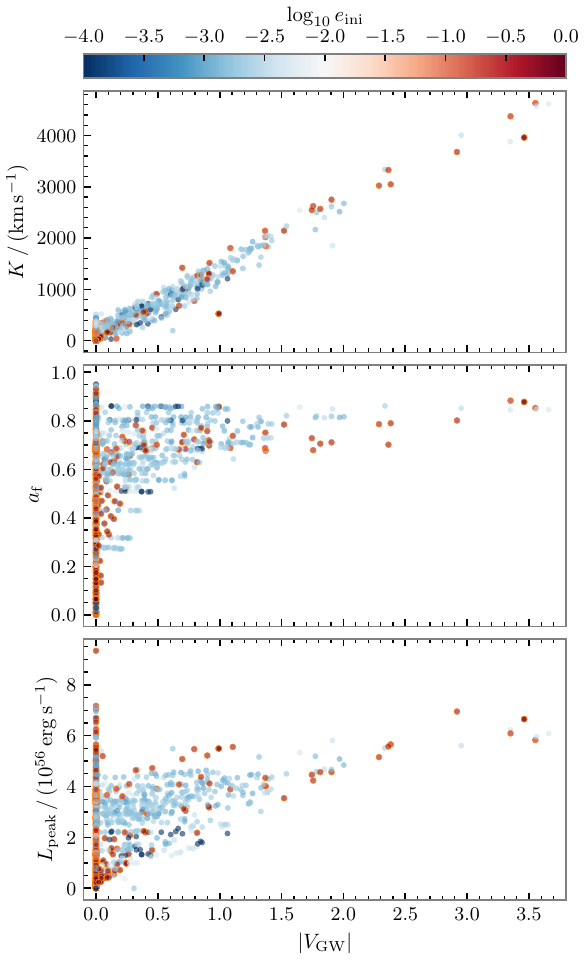}
    \caption{{\bf Relation of $\vgw$ and initial eccentricity} 
        From top to bottom, the three panels plot $\abs{\vgw}$ against the magnitudes of final kick, final spin and peak luminosity, respectively.
        Each point corresponds to one NR simulation in the RIT catalog, and is colored by the initial eccentricity (in log-scale); points with orange outlines are explicitly stated as eccentric in the catalog.
        We note that generally eccentric binaries tend to have mildly higher kick and luminosity, as well as $\vgw$.
    }
    \label{fig:VGW_spin_kick}
\end{figure}

In Fig.~\ref{fig:VGW_spin_kick}, from top to bottom, we have plotted the kick and spin magnitudes, together with the peak luminosity, against $\abs{\vgw} $, for simulation in the RIT catalog. 
Each point is coloured according to the initial eccentricity of the simulation, as reported in the RIT catalogue. 
The top panel shows that, as expected, eccentric binaries dominate the very large kick regions, especially when the resultant kick magnitude is above \SI{3000}{\km\per\s}, also corresponding to the largest values of $\vgw$.

The middle panel shows that the most extreme values of $\abs{\vgw}$ occur for binaries that result in spins around 0.6-0.8.
We note that the spin distribution is apparently capped by $\sim 0.9$, which is a limitation of the available simulation, which shows low $\vgw$ values. This is consistent with the fact that near-extremal BHs can only be produced with peculiar configurations with nearly aligned-spins, which necessarily have very small $\vgw$.

Finally, the bottom panel shows the peak luminosity as a measure of energy flux, which is directly correlated with the amplitude of the momentum flux, \ie~the kick. As expected from their strong precession and eccentricity, systems with large \vgw display larger luminosities, which implies that they should be more easily detected. We note, however, that current searches for BBHs ignore orbital precession~\cite{Harry:2016ijz}, damaging their sensitivity \cite{Bustillo:2016gid,Chandra2020_Nuria}.

\section{Regarding final spin of remnant BHs~\label{app:final_spin}}
\begin{figure}[htbp!]
    \centering
    \includegraphics[width=0.49\textwidth]{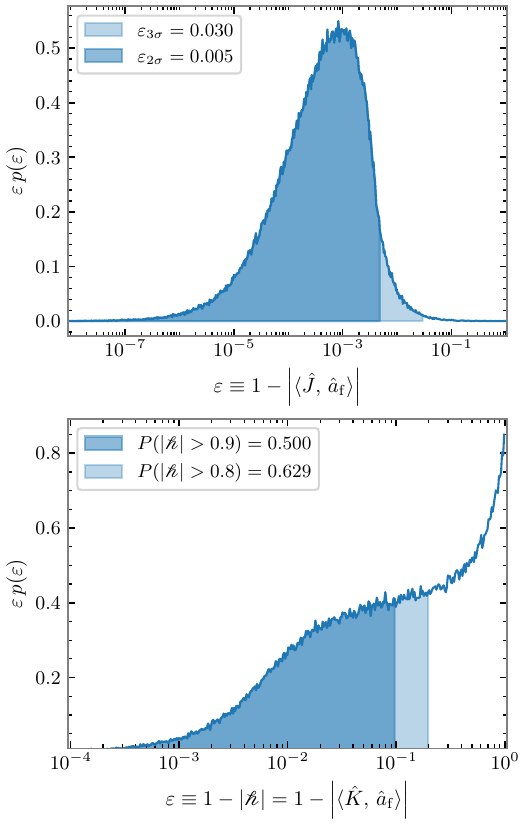}
    \caption{{\bf General alignments of the final spin of remnant black holes} 
        The top and bottom panels respectively show the distribution of the alignment of the final spin of remnant black holes with the total angular momentum $\vu*J$ and $\vu*K$.
        Similar to Fig.~\ref{fig:linear_relation}, these are computed using \num{500000} BBH using the \nrsur~waveform model.
        In the top panel, we represent the alignment of $\vu*J$ and $\vu*a_{\rm f}$ by the deviation of their inner product from 1, in log-scale. 
        The dark- and light-shaded areas denote 95.45\% ($2\sigma$) and 99.73\% ($3\sigma$) of all BBHs, respectively.
        The bottom panel shows the distribution of (absolute value of) helicities of these BBHs, and the shaded region is where $\abs{\mathscr{h}} > 0.9 $, which happens to be about 50\% of the ensemble.
    } 
    \label{fig:final_spins}
\end{figure}

In this last appendix, we will show empirically that the total angular momentum $\vb*J$ of of generic BBHs tends to align with the final spin of the remnant. 
The top panel of Fig.~\ref{fig:final_spins}, shows $\< \vu*J,\,\vu*a_{\rm f}>$ and its deviation from unity, for the \num{500000} BBHs with spins isotropically distributed we used in Fig.~\ref{fig:linear_relation}.
We represent its distribution in log-scale to highlight the strong alignment of these two vectors. In particular, for 99.73\% of all \num{500000} BBHs, the angle between $\vu*J$ and  $\vu*a_{\rm f}$ are well-below 0.247\,rad.
Furthermore, for 95.45\% ($2\sigma$) of them, this angle is less than 0.0996\,rad.
This should justify the interchangeability of $\vu*J$ and $\vu*a_{\rm f}$ in the results we show in the main text.
We highlight that among all \num{500000} BBHs that we have considered, only seven of them have $\< \vu*J,\,\vu*a_{\rm f}> < 0$.

Finally, regarding our notion of helicity, all the plots in the main text show the value of the helicity scaled by the kick magnitude, which concealed the fact that in most remnant BHs, this quantity clusters at $\pm 1$, as shown in the lower panel of Fig.~\ref{fig:final_spins}. In the plot, we see that under our isotropic spin distribution assumption, about half of all BBHs should have $\abs{\mathscr{h}} > 0.9 $, and about 62.5\% of them will have $\abs{\mathscr{h}} > 0.8 $.
This property shall be another unique feature of a population of BBHs with isotropic spins. For instance, for aligned spin BBHs, one expects the kick to be orthogonal with $\vu*J$ (and $\vu*a_{\rm f}$), which would yield null helicity. For BHs that are formed from other astrophysical processes, such as supernovae, perhaps the distribution of the helicity would be different as well. This suggests that mirror asymmetry in the population of BHs in the Universe could arise if BH formation was predominated by certain channels which have characteristic helicity distributions. 

\end{appendix}

\bibliography{bibliography}
\end{document}